\documentclass{jfm}

\usepackage{graphicx}
\usepackage{newtxtext}
\usepackage{newtxmath}
\usepackage{natbib}
\usepackage{hyperref}
\usepackage{booktabs}
\usepackage{float}
\usepackage{placeins}
\usepackage{caption}
\usepackage{makecell}
\usepackage{subcaption}
\usepackage{caption}
\usepackage{capt-of}
\usepackage{ragged2e}

\usepackage{pdflscape}
\usepackage{siunitx}
\hypersetup{
    colorlinks = true,
    urlcolor   = blue,
    citecolor  = black,
}

\newcommand{\RomanNumeralCaps}[1]
\linenumbers

\title{Transition to turbulence in swirl created by a vertical magnetic field in liquid metal electrodes}

\author{Uma Siva Sankar Kompalli\aff{1}
  \corresp{\email{}},
  Avishek Ranjan\aff{2}
 \and Suneet Singh\aff{2}}

\affiliation{\aff{1}Department of Energy Science and Engineering, Indian Institute of Technology Bombay, Powai, Mumbai, 400076, India
\aff{2}Department of Mechanical Engineering, Indian Institute of Technology Bombay, Powai, Mumbai, 400076, India}

\begin{document}
\maketitle

\begin{abstract}
Flow transitions in electromagnetically-driven swirl in a cylindrical domain are investigated using three-dimensional simulations for a liquid metal battery electrode. The parameter space considered: applied current $I=11.11-100A$; Current density, $J_0 \approx 4,000-35,000 A/m^2$, domain radius, $R=3-9cm$, domain height, $H=3$ and $4.5cm$). The background magnetic field, $B_0$, strength is varied systematically from 0.1 to 20 G to characterize the route to chaos. \cite{davidson2022swirling} found the onset of chaos occurs around $1-1.5 G$ for aspect ratio  $1/6$ and current of 100A, however, this is shifted to $16-17 G$ for a lower current of 16A and aspect ratio $1/2$. 
Across a range of geometrical configurations and forcing conditions, including variations in aspect ratio, current and current density, a consistent sequence of dynamical transitions is evident from phase portraits, FFT spectrum, 0-1 chaos test, autocorrelation function and correlation dimension. The flow evolution is by the Ruelle--Takens-Newhouse route to chaos, with the strange non-chaotic attractor (SNA) regime providing an intermediate state prior to the onset of chaos. In our results, it is observed that, for a fixed aspect ratio, the transition is governed by a dimensionless group, $Q_m$, representing the ratio of applied Lorentz force to viscous force. The critical value of $Q_m$ at the onset of oscillatory flow is found to be nearly invariant except when the domain height is varied. Moreover, our results follow the $Re \sim Q_m^{2/3}$ scaling law expected from torque balance moderately-forced electro-magnetically driven swirl flows.

\end{abstract}



\section{Introduction}

\begin{figure}
\centering

\begin{subfigure}{0.46\textwidth}
    \centering
    \includegraphics[
    width=\linewidth,
    trim=13cm 8cm 7cm 6cm,
    clip
    ]{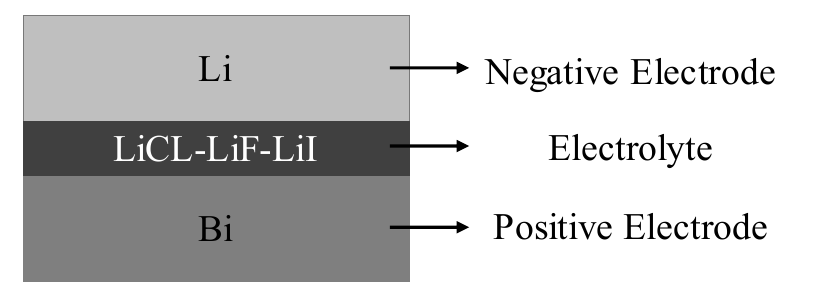}

    \phantomcaption
    \label{fig:LMB_a}
\end{subfigure}
\hfill
\begin{subfigure}{0.46\textwidth}
    \centering
    \includegraphics[
    width=\linewidth,
    trim=12cm 8cm 8cm 6cm,
    clip
    ]{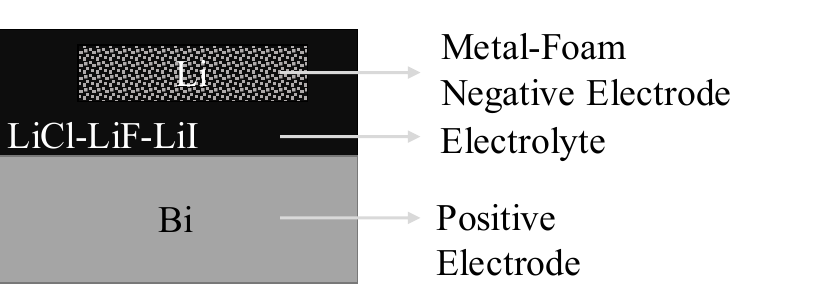}

    \phantomcaption
    \label{fig:LMB_b}
\end{subfigure}
\captionsetup{width=\textwidth}
\caption{
Schematic of liquid metal battery configurations:
(a) density-stratified model and
(b) metal-foam configuration.
}
\label{fig:LMB}

\end{figure}

Liquid metal batteries (LMBs) have emerged as a promising technology for grid-scale electrical energy storage for stationary applications due to their projected low cost, safety at higher currents and temperature, and long cycle life \citep{kim2013liquid, bradwell2012magnesium,kelley2018fluid}. In a typical LMB, liquid metal electrodes separated by a molten salt electrolyte, forming a stratified three-layer configuration as shown in figure \ref{fig:LMB_a}. However, in a realistic LMB, the negative electrode is kept in a metal foam (figure\ref{fig:LMB_b})  \cite{Ning2015,Ouchi2016Patent}. The performance and safety characteristics of LMBs are reported to be influenced by the fluid motion which occurs in the presence of thermal (\cite{Shen2016,Alberto2017,Thomas2017}), solutal (\cite{Personnettaz2019,Herreman2021,Duczek2023}), interfacial and electromagnetic effects (\citep{Weber2017, horstmann2018coupling}). It is well known in the MHD literature, that the swirling fluid flow due to an external (vertical) magnetic field dominates over the effect of current's own (azimuthal) magnetic field( \cite{Davidson1999,Frick2022}). In their study carried out for a Li-Bi LMB with current capacity of 100A, using an exact solution of the swirling Lorentz force, \citet{davidson2022swirling} showed that an external magnetic field as small as 0.5G can lead to oscillatory flow and, in fact, a field of 1.5-2G can make the flow turbulent (for comparison, a fridge magnet has a field strength of $\approx 50-100G$). Vigorous turbulence may eventually lead to growing interfaces ultimately leading to a short-circuit. However, the mechanism of transition to turbulence at such a small field was not fully investigated. Moreover, this finding needs to be further examined at other current values and at other aspect ratios, which is the focus of the present study.

\begin{figure}
\centering
    \includegraphics[
    width=\linewidth,trim=1cm 2cm 3cm 3cm,
    clip
    ]{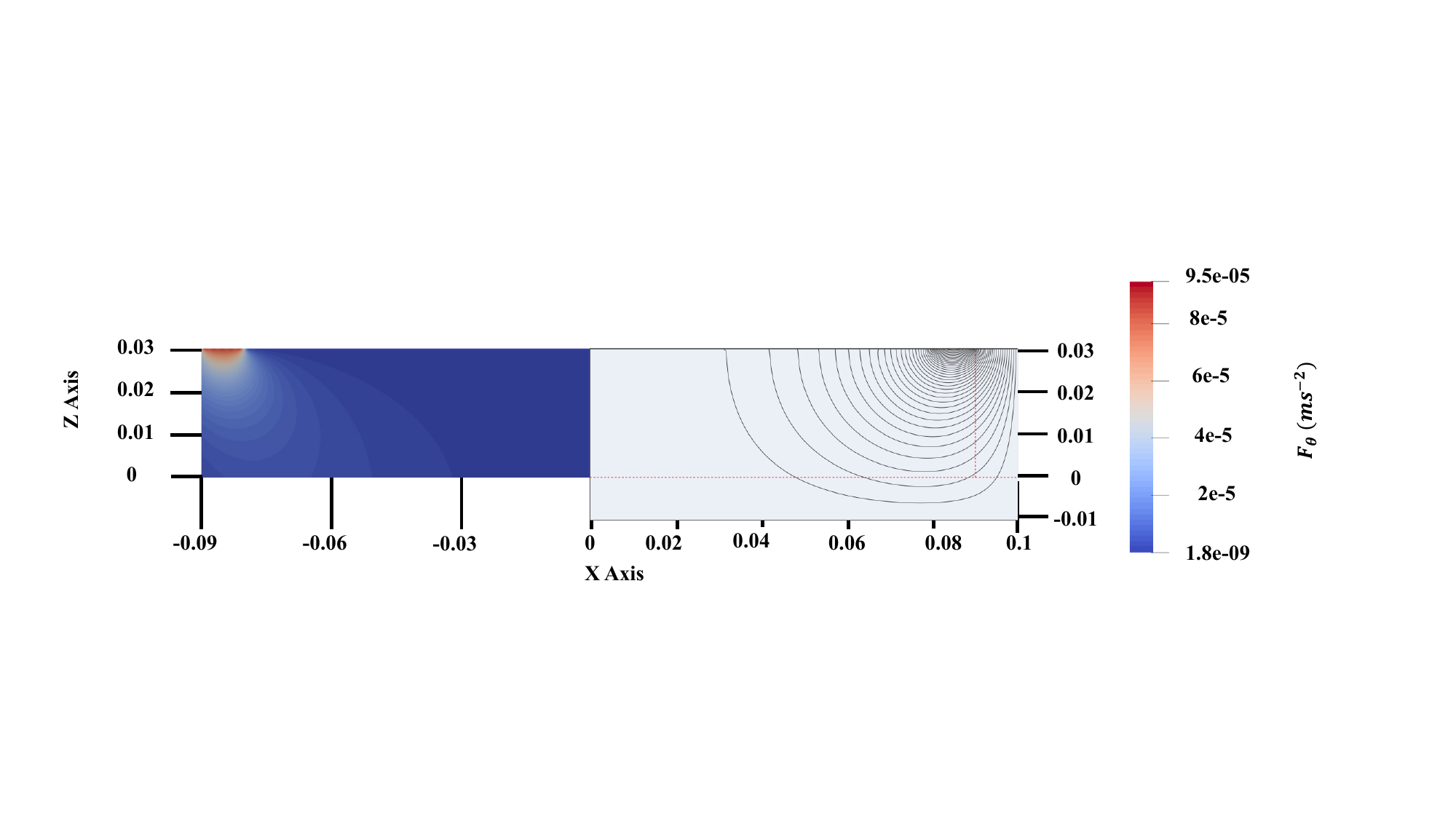}
\captionsetup{width=\textwidth}
\caption{
Lorentz force distribution in the domain and current lines in the domain along with the container when $B=0.27G$ is applied for aspect ratio $1/6$ for case 1
}
\label{fig:F_theta}

\end{figure}

From a swirling flow perspective, it is expected that variations in forcing and geometry parameters, such as current density, magnetic field and aspect ratio \textit{can} influence the onset of instabilities and the flow behaviour. This can be attributed to an altered balance between the kinetic energy supplied (imparted by the Lorentz force near the point of current entry/exit) and that is dissipated through viscous action near the walls. Theoretical frameworks for understanding the onset of centrifugal instability in swirling flows and MHD instability are well established\citep{chandrasekhar1961hydrodynamic, davidson2017}. However, in most analytical studies, the boundary conditions are typically uniform, i.e. Dirichlet or Neumann type. In a configuration with complex forcing functions that can create swirling flow in LMBs (see the current lines in figure\ref{fig:F_theta} and the Lorentz force contours), the transition can be most readily studied using numerical simulations. Moreover, most existing studies have focused on the onset of instability or steady-state behavior, with limited attention given to the full sequence of transitions that occur as the forcing amplitude is increased. A comprehensive understanding of these transitions is therefore necessary for predicting and controlling flow behavior in LMBs or similar configurations.

The transition from steady flow to turbulence in other fluid systems often proceeds through a sequence of bifurcations involving intermediate dynamical states. Classical studies have shown that such transitions may follow the Ruelle--Takens--Newhouse scenario, in which a steady flow loses stability via a Hopf bifurcation, followed by the emergence of additional frequencies and eventual transition to chaos \citep{ruelle1971nature,newhouse1978occurrence}. Similar transition pathways have been reported in other fluid mechanical systems, such as differentially heated cavities \citep{lequere1998chaos} and thermo-acoustic flows \citep{kashinath2014thermoacoustic}, where complex intermediate regimes arise prior to fully developed chaos. These studies highlight the importance of analyzing time-dependent behavior to understand the underlying nonlinear dynamics.

Despite these advances, the route to chaos in electromagnetically driven flows, relevant to LMBs and other similar configurations, remains largely unexplored. In particular, the existence and role of intermediate regimes, such as multi-harmonic periodic states or strange non-chaotic attractors, have not been systematically investigated in this context. Moreover, while it is known that the ratio of Lorentz force to the viscous force is important, the dimensionless parameter that controls this behavior is not the Hartmann number $Ha=B_0L\sqrt\frac{\sigma}{\rho \nu}{ }$, where $L, \rho, \nu, \sigma$ are length scale, fluid density, kinematic viscosity and electrical conductivity,  which is based on the applied field alone and excludes the applied current. As evident later from our results, the appropriate dimensionless number that governs the dynamical behavior is $Q_m =J_0B_0R^3/\rho\nu^2$, hereafter referred to as the modifed Chandrasekhar number. Dimensional scaling laws in terms of $U$ vs $J_0$ or $B_0$, proposed earlier by \citep{Herreman2021}, can be combined in terms of $Re, Q_m$ for inductionless regimes.

In the present study, therefore, we investigate the transition to unsteady and chaotic flow in a cylindrical liquid metal domain subjected to an imposed vertical magnetic field and an applied current. Using extensive three-dimensional direct numerical simulations, the background magnetic field strength is varied systematically to characterize the sequence of dynamical transitions. An extensive range of configurations is considered, including variations in aspect ratio, current and current density ($I=11.11-100A$; $J_0 \approx 4,000-35,000 A/m^2$, $R=3-9cms$,  $H=3$ and $4.5cms$).

The objectives of this work are threefold. First, to identify and characterize the sequence of dynamical regimes, starting from steady flow to chaos in electromagnetically driven liquid metal systems. Second, to examine the robustness of this transition across different geometrical and forcing conditions. Third, to establish a unifying framework for the transition based on a modified Chandrasekhar number, representing the ratio of (applied) Lorentz to viscous forces. In section \ref{section2} the mathematical formualtion of the model along with numerical methodology considered was presented. In section \ref{section3} the influence of control parameters, $J$, $B_0$, $R$, $H$ and aspect ratio on onset of Hopf point and the subsequent route to chaos. A new non-dimensional number, which is the ratio of Lorentz force to the viscous force, was proposed . The results were compared with existing scaling laws, and a non-dimensional framework for expressing the same results was presented. Finally, section \ref{section4} summarizes the findings and concludes the paper.

\section{Mathematical formulation and numerical methodology}\label{section2}

\begin{figure}
\centering
    \includegraphics[
    width=\linewidth
    ]{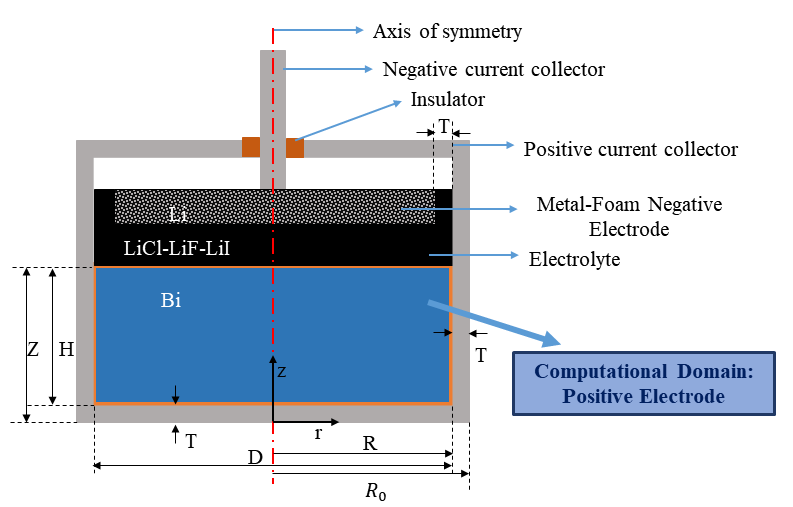}
\captionsetup{width=\textwidth}
\caption{
Realistic schematic of Li-Bi liquid metal battery with metal-foam configuration
}
\label{fig:LMB_r}

\end{figure}

A cylindrical lithium--bismuth liquid metal battery (LMB), corresponding exactly to the configuration considered by \citet{davidson2022swirling}, is investigated, as illustrated in figure~\ref{fig:LMB_r}. The battery consists of a liquid bismuth positive electrode beneath a molten salt electrolyte and a lithium-containing porous negative electrode. An electric current is supplied from the top current collector, passes through the porous electrode and electrolyte, and enters the liquid bismuth electrode.

The thermophysical properties of liquid bismuth at \SI{500}{\celsius} are adopted from \citet{kelley2018fluid,Weber2020}. The density, kinematic viscosity, and electrical conductivity are taken as $\rho = 9720~\mathrm{kg\,m^{-3}}$, $\nu = 1.2 \times 10^{-7}~\mathrm{m^2\,s^{-1}}$, and $\sigma = 7.3 \times 10^{5}~\mathrm{S\,m^{-1}}$, respectively. Unless otherwise stated, the gap between negative electrode and the positive current collector is fixed at $T = 1~\mathrm{cm}$ for all cases.

For simplicity, a no-slip boundary condition is imposed at the top boundary of the bismuth layer. The sidewalls and bottom boundary are also treated as no-slip surfaces. This assumption allows the focus to remain on the bulk flow dynamics within the positive electrode.

The flow is driven by the interaction between the imposed electric current, $\mathbf{J}$ and an externally applied vertical magnetic field,
\begin{equation}
\mathbf{B} = B_0 \,\hat{\mathbf{e}}_z,
\end{equation}
to generate azimuthal motion through the Lorentz force. The resulting flow is primarily governed by the azimuthal component of the Lorentz force, which induces swirling within the liquid metal.

To calculate $\mathbf{J}$, potential has to be known, since  $\mathbf{J}=-\sigma\nabla\phi$, where $\sigma$ is the electrical conductivity and the $\phi$ is potential. Since $\mathbf{J}$ is solenoidal, $\phi$ is harmonic i.e.,
\begin{equation}
    \nabla^2\phi=0
    \label{eq:harmonic}
\end{equation}
As $\mathbf{J}$ is axisymmetric, it can be expressed in Stokes stream function, $\mathbf{\Psi}(r,z)$. Then combined equation can be expressed as follows:
\begin{equation}
    \mathbf{J}=\nabla\times[(\mathbf{\Psi}/r)\hat{\mathbf{e}}_\theta]=-\sigma\nabla\phi
    \label{eq:stokesohm}
\end{equation}
Equation~\eqref{eq:harmonic} is solved over the cylindrical domain $0 \le r \le R_0$, $0 \le z \le Z$, comprising the bismuth layer and the surrounding container wall (see figure~\ref{fig:LMB_r}), subject to the following boundary conditions. At the base of the container ($z=0$) and along the outer wall ($r=R_0$), no current crosses the boundary, so
\begin{equation}
    \mathbf{J}\cdot\hat{\mathbf{n}} = 0 \quad \Longleftrightarrow \quad \frac{\partial \phi}{\partial n} = 0.
\end{equation}
At the upper boundary ($z=Z$), the potential is prescribed directly from the current-injection, since the electrolyte is far more resistive than the bismuth, the current path from the base of the lithium electrode to the bismuth surface is essentially vertical, giving a uniform potential $\phi = V$ over the region below the lithium electrode ($r < R-T$); the potential is taken as $\phi = 0$ within the container wall ($R < r \le R_0$); and a linear variation, $\phi = (R-r)V/T$, is assumed across the annular electrolyte-filled gap in between ($R-T < r < R$), where $T$ is the gap width. These conditions complete the boundary conditions needed to solve \ref{eq:stokesohm} for $\phi$, from which $\mathbf{J}$ and the corresponding azimuthal Lorentz force distribution are obtained.

Although the applied current generates an induced magnetic field, its influence on the flow is negligible compared to the externally imposed magnetic field. This is justified by the criterion that when $|B_z| \gg 0.01 B_\theta$, the Lorentz force is dominated by the interaction of the current with the imposed vertical field \cite{Davidson1999, Frick2022}. In the present study, the imposed magnetic field is chosen such that this condition is satisfied, ensuring that the dominant forcing arises from the azimuthal Lorentz force.

Under these assumptions, the flow is governed by the incompressible Navier--Stokes equations with the Lorentz force term, $\mathbf{F_L}=\mathbf{J}\times\mathbf{B}$
\begin{align}
\nabla \cdot \mathbf{u} &= 0, \\
\rho \left( \frac{\partial \mathbf{u}}{\partial t} + \mathbf{u} \cdot \nabla \mathbf{u} \right) &= -\nabla p + \mu \nabla^2 \mathbf{u} + \mathbf{F_L},
\end{align}
where $\mathbf{u}$ is the velocity, $p$ is the pressure, $\rho$ is the density, $\mu$ is the dynamic viscosity and $\mathbf{J}$ is the imposed current density.The induces current is ignored as the maximum magnetic Reynolds number, ${Re_m}=UR\mu_0 \sigma$, is $\approx0.01$, where $U$ is characteristic flow velocity, $\mu_0$ is the magnetic permeability of free space.

\FloatBarrier

\subsection{Lorentz force formulation}

The flow is driven by the azimuthal Lorentz force arising from the interaction between the imposed current density and the externally applied vertical magnetic field. The current distribution is obtained from the electric potential formulation, as described in \citet{davidson2022swirling}.

The  Lorentz force per unit volume, used in the momentum equations, is given by
\begin{equation}
F_\theta(r,z)
=
\frac{2\sigma V B_0}{\rho}\frac{R}{R_0^2}
\sum_{n}
\frac{\tilde{J}_1(n)\, J_1\!\left(\delta_n r/R_0\right)}{J_0^2(\delta_n)}
\frac{\cosh\!\left(\delta_n z/R_0\right)}{\cosh\!\left(\delta_n Z/R_0\right)},
\end{equation}
where $J_0$ and $J_1$ are Bessel functions of the first kind, $\delta_n$ are the corresponding eigenvalues, and $\tilde{J}_1(n)$ is defined through the radial current distribution. (A detailed derivation of $F_\theta$ is given in Appendix \ref{appA} for completeness). For case 1, with an aspect ratio (H/D) of 1/6,  $F_\theta$ is shown in figure \ref{fig:F_theta} on the XZ plane.

\subsection{Numerical simulation methodology}

The governing equations are solved using the finite-volume method with a PISO algorithm for pressure--velocity coupling using OpenFOAM. The advection and gradient terms are discretized using a second-order Gauss linear scheme, while the diffusion term is discretized using a Gauss linear orthogonal scheme. Temporal discretization is performed using a Crank-Nicolson scheme.

A systematic grid and time-step sensitivity study is conducted to ensure numerical accuracy while maintaining computational efficiency. Several grids ranging from $1.2 \times 10^3$ to $1.6 \times 10^6$ control volumes, together with time steps between $0.25$ and $0.0025$, are examined, out of these some are shown in table \ref{tab:grid_independence}. The temporally averaged velocity at a representative location in the domain is monitored to assess numerical convergence. Third configuration was considered after considering computational cost and accuracy. Overall, a total of 592 three-dimensional simulations were performed for parametric studies over different aspect ratios, applied currents, and background magnetic fields.

The results indicate that the variation in velocity between the finer grids is small, demonstrating grid convergence. Based on a balance between accuracy and computational cost, the grid with approximately $1.3 \times 10^6$ cells is adopted for all simulations. A time step of $0.005$ is found to provide a good compromise between temporal accuracy and computational efficiency, and is therefore used throughout the study.

\FloatBarrier
\begin{center}
\captionsetup{width=\textwidth}
\captionof{table}{Grid independence test: Time-averaged velocity magnitude at point $(0.08748,\,0,\,0.0069)$ for Case 1, aspect ratio $1/6$ and $B_0 = 1.5\,\mathrm{G}$}
\label{tab:grid_independence}
\begin{tabular}{cccc}
\toprule
Case no. & Number of cells & Time step & $\langle U \rangle$ (cm/s) \\
\midrule
1 & 688704  & 0.005 & 0.658 \\
2 & 999020  & 0.005 & 0.675   \\
3 & 1325520 & 0.005 & 0.709  \\
4 & 1541232 & 0.005 & 0.731 \\
\bottomrule
\end{tabular}
\end{center}

\vspace{\baselineskip}

The present numerical framework follows that of \citet{davidson2022swirling}, and the same solver  is employed. Additional details about the  validation are mentioned therein. Specifically, the simulated azimuthal velocity, $U_\theta$, along the radial direction is compared with theoretical prediction and power law profile, while the radial variation of the azimuthal force, $F_\theta$, is compared with the corresponding power-law relationship.

\section{Route to chaos}\label{section3}

The transition to unsteady and chaotic flow is investigated through four sets of simulations designed to isolate the effects of geometry and forcing. In the first case, the aspect ratio is varied by changing the radius while keeping the total current and height fixed. In the second case, the current is varied for fixed aspect ratio, thereby altering the current density. In the third case, the radius and current are varied simultaneously such that the current density and height remain constant. Finally, in the fourth case, the aspect ratio is varied by changing only H while maintaining a fixed current.

For each configuration, the background magnetic field strength is systematically varied over the range $B_0 = 0.025$--$20\,\mathrm{G}$ in order to characterize the sequence of dynamical transitions. This allows the identification of successive bifurcations and the corresponding critical values of $B_0$ associated with the onset of unsteadiness and the transition to chaos.

The flow behavior is analyzed using time series extracted at selected locations within the domain. Based on observations across multiple probe points, a representative location at $r/R = 0.97$ and $z/H = 0.23$, where oscillations are observed to first emerge, is chosen for detailed analysis. The resulting sequence of flow regimes is presented in the following sections.

\subsection{Case 1: Effect of Radius, Aspect Ratio and current density}
\FloatBarrier
\subsubsection{Steady State Regime}
\begin{figure}
\centering

\includegraphics[
width=1.1\textwidth,
keepaspectratio
]{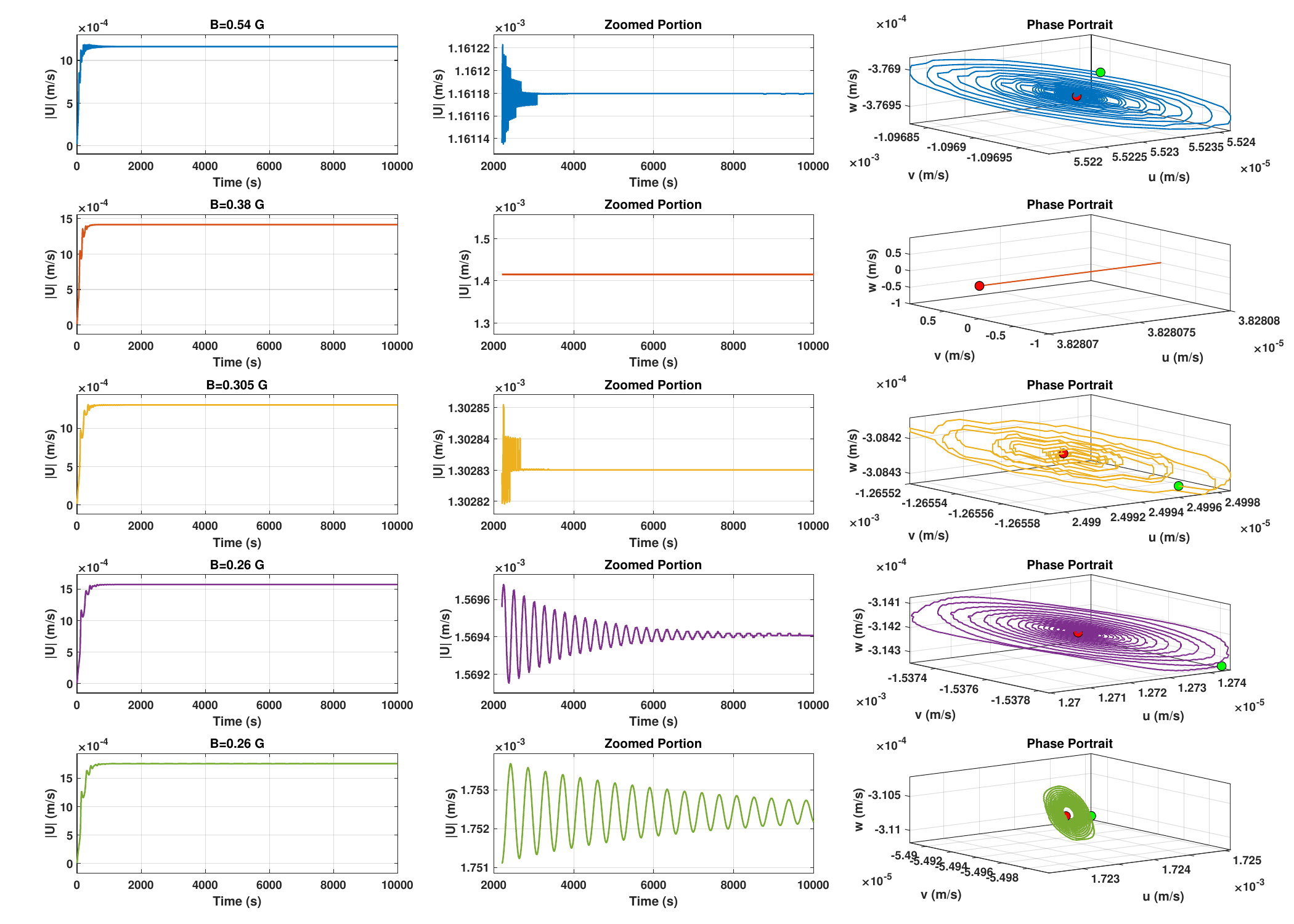}
\captionsetup{width=\textwidth}
\caption{
Steady-state regime for case 1 showing decaying oscillations
and stable phase portraits at different aspect ratios.
}
\label{fig:c1steadystate}

\end{figure}

For all aspect ratios, the initial flow in the bismuth layer is steady and laminar, indicating that the Lorentz force balances viscous forces and pressure gradients. The range over which the flow remains steady varies with aspect ratio, as shown in figure ~\ref{fig:c1steadystate}. \\
Figure~\ref{fig:c1steadystate} shows the presence of decaying oscillations in the flow. This behavior suggests that the system is close to a critical point or bifurcation threshold, where small perturbations may lead to qualitative changes in the flow dynamics. Such oscillations indicate a transitional state in which the system approaches the onset of instability. The phase portraits in figure ~\ref{fig:c1steadystate} exhibit an inward spiral which reflects damped dynamics, confirming that the flow remains stable.

\begin{figure}
\centering

\includegraphics[
width=1.1\textwidth,
keepaspectratio
]{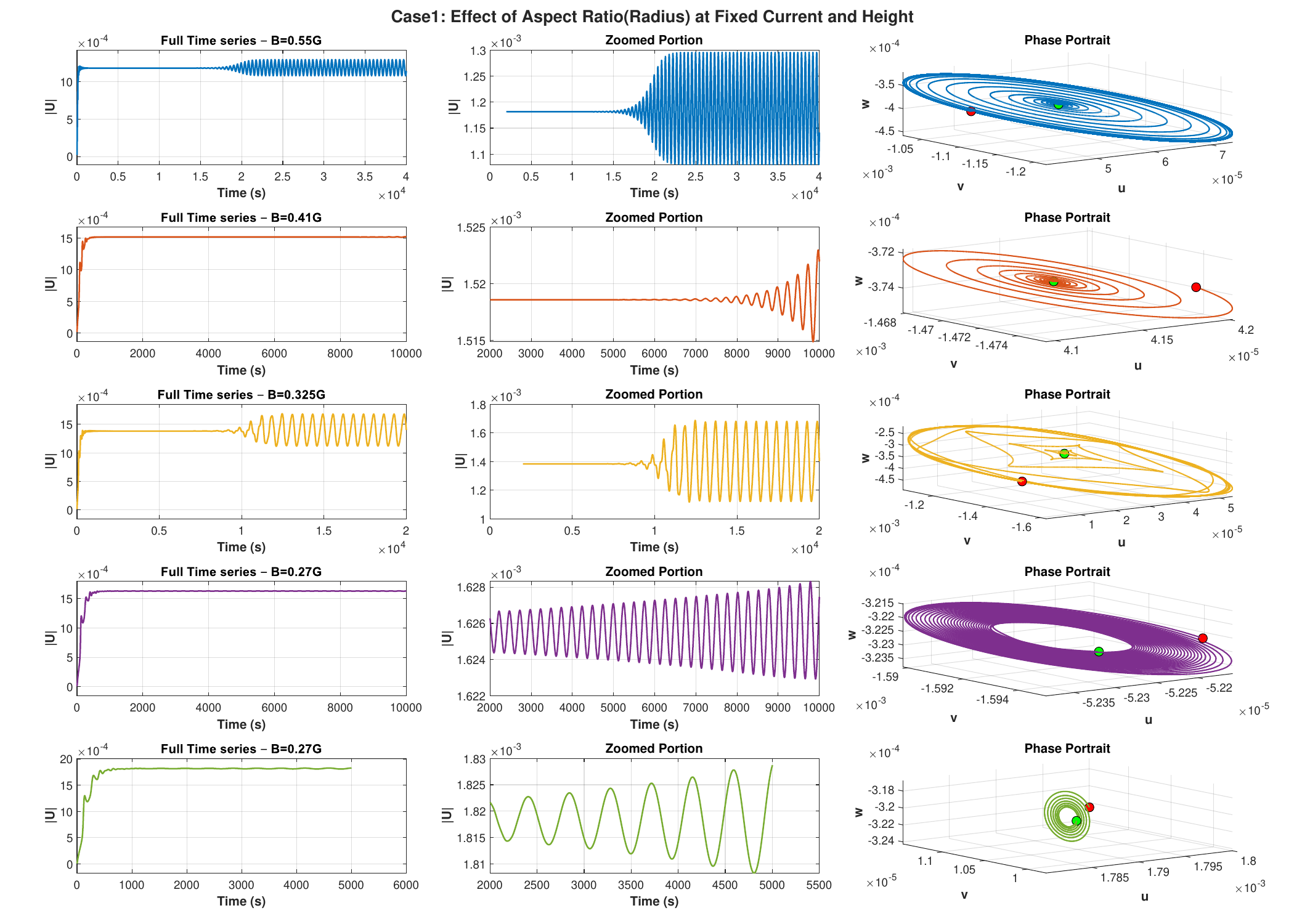}
\captionsetup{width=\textwidth}
\caption{
Unstable regime for case 1 showing growing oscillations
leading to a limit cycle at different aspect ratios.
}
\label{fig:c1growingoscillations}

\end{figure}

\subsubsection{Hopf Point}
With increasing background vertical magnetic field (varies for different aspect ratios), oscillations begin to appear and grow, as shown in figure ~\ref{fig:c1growingoscillations}, indicating the onset of a Hopf bifurcation. In several cases, both decaying and  growing oscillations are observed, allowing the Hopf point to be estimated by interpolation. The range over which this bifurcation occurs is summarized in table ~\ref{tab:combinedCR}.

However, in configurations with smaller radius (i.e., higher aspect ratio), the oscillations are strongly damped and the growth rate of the unstable mode is very small. Consequently, the system exhibits long transient behavior that appears nearly steady before oscillations eventually develop. In such cases, simulations were run longer to determine whether sustained oscillations occur, as illustrated in figure ~\ref{fig:c1growingoscillations}.

This delayed onset of oscillations is consistent with observations in other thermo-fluid systems. For example, \citet{lequere1998chaos} reported long transient states near the onset of unsteadiness in differentially heated cavity configurations, while \citet{kashinath2014thermoacoustic} showed that trajectories may temporarily approach unstable states before evolving towards the final attractor. As a result, the flow may appear steady over intermediate time intervals before the oscillatory limit cycle is established. This indicates the emergence of an oscillatory instability driven by the competition between swirl due to Lorentz force and damping at the walls due to viscous forces.

The instantaneous flow patterns after the Hopf point in the first case are shown in the first row of figure ~\ref{fig:UPol_C1}, where poloidal velocities are shown by slicing the geometry on the XZ plane. Across all aspect ratios considered in this regime, the flow remains axisymmetric, with a well-defined secondary circulation clearly visible. Although the maximum velocity occurs near the top and bottom boundaries, the onset of oscillations is observed at approximately two-ninths of the domain height, measured from the bottom.

\subsubsection{Periodic Regime with single frequency}
In the vicinity of the critical point, the growing oscillations do not settle to a well-defined amplitude over the same time interval. However, as $B_0$ increases further (varies for different aspect ratios), the oscillations stabilize to a finite amplitude and a clear limit cycle emerges, as shown in the second rows of the phase portraits in figures ~\ref{fig:c1AR12}--\ref{fig:c1AR16}. Spectral analysis using the fast Fourier transform (FFT) reveals a single dominant frequency close to the critical point. As the background vertical magnetic field increases, a second frequency becomes apparent.

The instantaneous flow patterns for this regime in the first case, are shown in the second row of figure ~\ref{fig:UPol_C1}. In this regime, as the domain becomes deeper (i.e., with increasing aspect ratio), the primary and secondary flow structures undergo significant changes and become the dominant feature of the overall flow.

\subsubsection{Periodic Regime with Harmonics }
With further increase in $B_0$ (varies for different aspect ratios), the solution becomes oscillatory and a stable limit cycle is observed. However, three distinct frequencies appear, as shown in the third rows of figures ~\ref{fig:c1AR12}--\ref{fig:c1AR16}, indicating the emergence of a multi harmonic periodic regime. The additional frequencies are harmonics of the fundamental frequency, indicating that the flow remains periodic, with nonlinear harmonic generation rather than quasi-periodic behavior.

In the present case, for some values of the background magnetic field between the Hopf point and the onset of the three-frequency state, a second frequency is also observed that is approximately twice the fundamental frequency. 

\begin{landscape}
\begin{figure*}
\captionsetup{justification=justified,singlelinecheck=false}
\centering
\includegraphics[
height=0.8\textwidth,
keepaspectratio,trim=0cm 0cm 0cm 1cm,
clip]{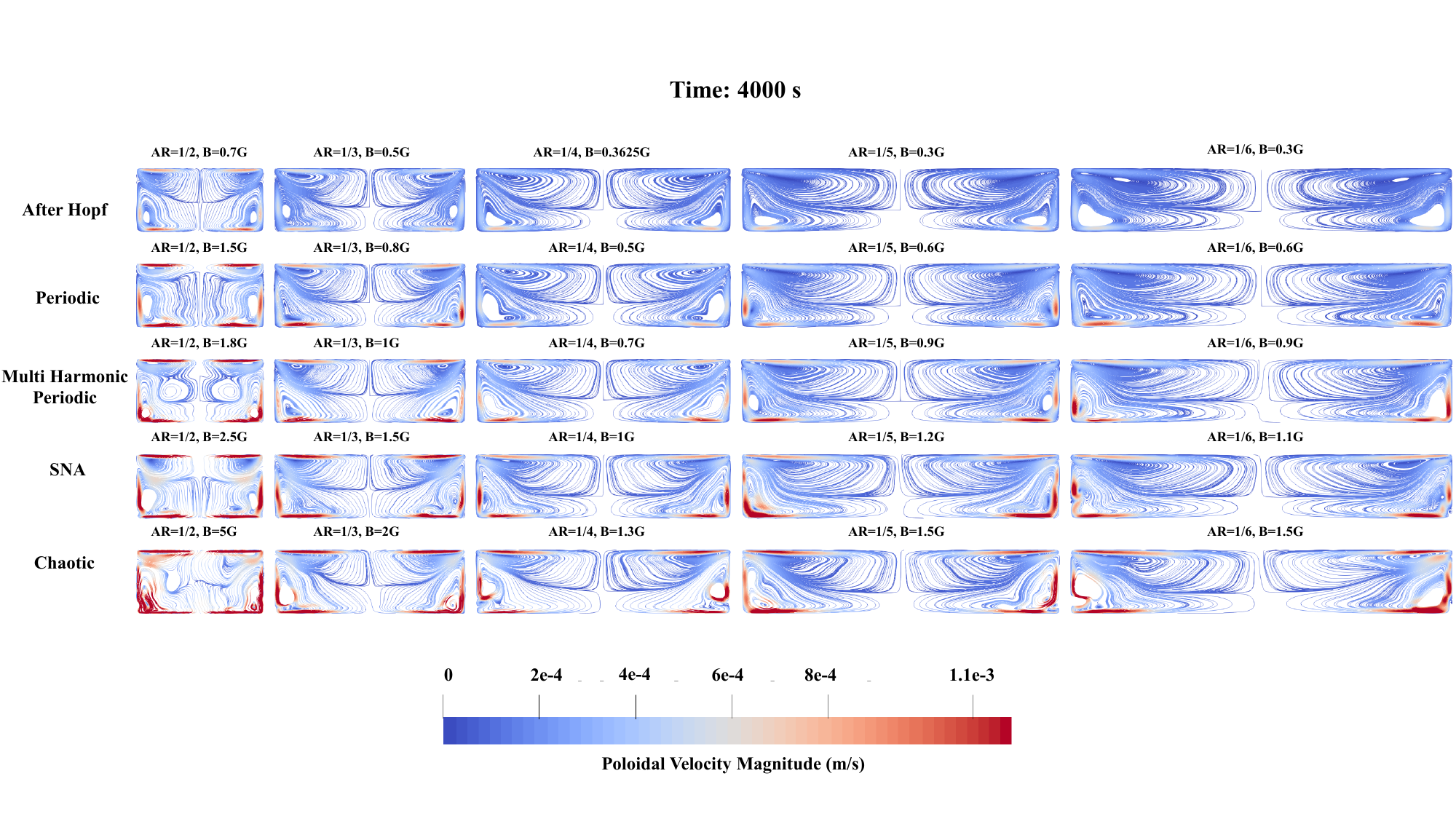}
\caption{Instantaneous Poloidal Velocity magnitude for Aspect Ratio 1/2, 1/3, 1/4, 1/5, 1/6 with applied current 100A at 4000 Seconds at different Background Magnetic fields for case1 }
\label{fig:UPol_C1}
\end{figure*}
\end{landscape}

The instantaneous flow patterns for this regime, in the first case, are shown in the third row of figure ~\ref{fig:UPol_C1}. At larger aspect ratios, a flow that resembles a ``recirculation bubble'' is observed near the centreline, similar to the vortex breakdown ``bubble'' reported by \cite{Escudier1984}. As the magnetic field strength increases, the azimuthal Lorentz force intensifies, driving a stronger swirling flow. The resulting adverse axial pressure gradient, induced by this swirl, decelerates the axial flow along the centreline. At a critical location, the axial velocity drops to zero, forming a stagnation point, before reversing direction further downstream and giving rise to a closed recirculation region. This behaviour is characteristic of vortex breakdown, as described by \cite{Escudier1984}. The present simulations exhibit a similar flow behaviour, suggesting that the strong electromagnetically driven swirl induces a vortex breakdown mechanism analogous to that observed in rotating endwall-driven flows.

\subsubsection{Strange Non-Chaotic Attractors}
With further increase in $B_0$ (varies for different aspect ratios), the flow becomes increasingly complex as it approaches a chaotic regime. However, spectral and statistical diagnostics, including the fast Fourier transform (FFT), autocorrelation function (ACF) and the 0-1 test for chaos (see Appendix~\ref{appB.1} for details ), indicate that the flow has not yet fully transitioned to chaos. The phase portraits exhibit closed but distorted trajectories, while the FFT shows only a limited number of discrete frequencies rather than a broadband spectrum.

The ACF decays gradually with lag, as shown in figures ~\ref{fig:c1AR12}--\ref{fig:c1AR16}, indicating persistent temporal correlations. Consistently, the 0-1 test yields non-chaotic behavior.

Taken together, these features—non-broadband spectra, slow decay of the ACF, non-chaotic 0-1 test results, and geometrically irregular yet bounded phase portraits—are characteristic of a strange non-chaotic attractor (SNA) \cite{Beeraiah_2024, Singh_Nair_2025}.

\begin{figure}
\centering
\includegraphics[width=\textwidth,keepaspectratio]{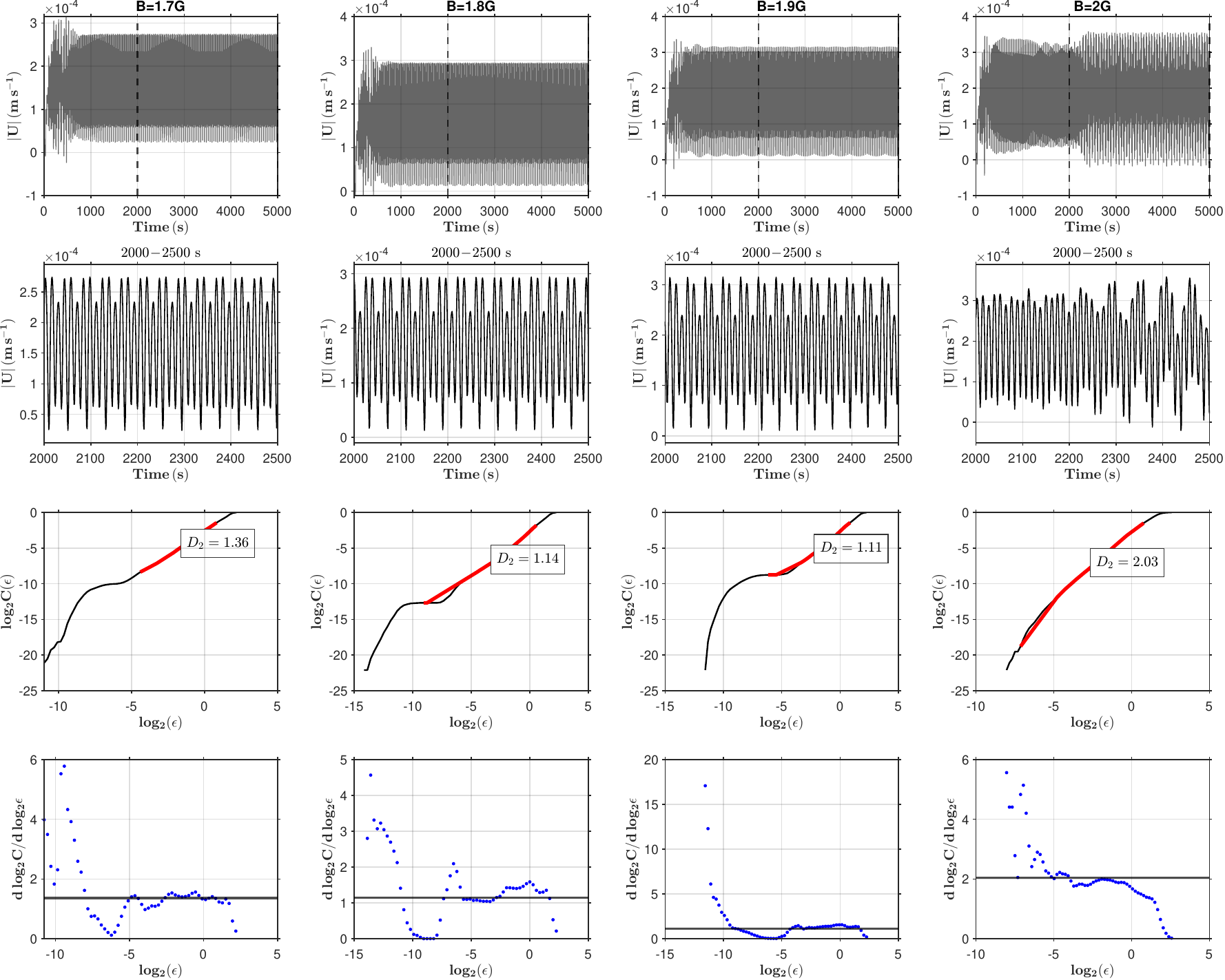}
\captionsetup{width=\textwidth}
\caption{Correlation dimension for Aspect ratio = 1/2 for case 1}
\label{fig:c1CD}
\end{figure}

To further assess the presence of an SNA,  the correlation dimension (see Appendix~\ref{appB.2} for  details) is employed to characterize the dynamics as illustrated in figure ~\ref{fig:c1CD}. This indicates strong nonlinear interactions and a loss of temporal coherence.

With the exception of the aspect ratio 1/2 case, all other cases retain the same flow structure observed in the multi-harmonic regime, albeit with increased velocity. For the aspect ratio 1/2 case, however, the ``recirculation bubble'' collapses, and the flow begins to lose its axisymmetry.

\subsubsection{Chaos}
\begin{landscape}
\begin{figure}
\centering
\includegraphics[
width=1.5\textwidth,
height=1.5\textheight,
keepaspectratio,trim=1cm 1.5cm 1.5cm 0.5cm,
    clip
]{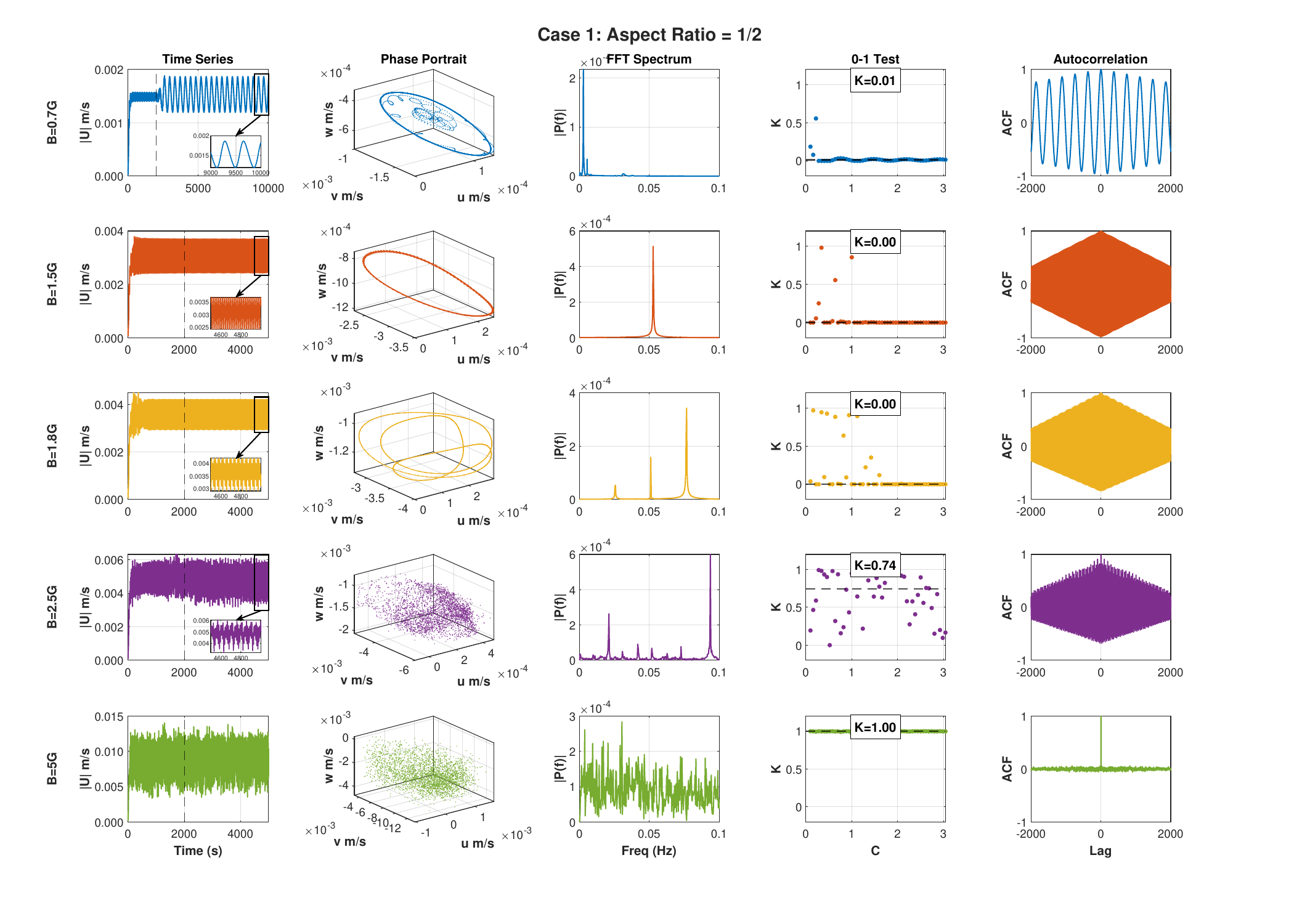}
\caption{Time series, phase portraits, FFT, 0-1 Test for chaos, Autocorrelation function for case1 for aspect ratio 1/2 at different B values}
\label{fig:c1AR12}
\end{figure}
\end{landscape}

\begin{landscape}
\begin{figure}
\centering
\includegraphics[
width=1.5\textwidth,
height=1.5\textheight,
keepaspectratio,trim=1cm 1.5cm 1.5cm 0.5cm,
    clip
]{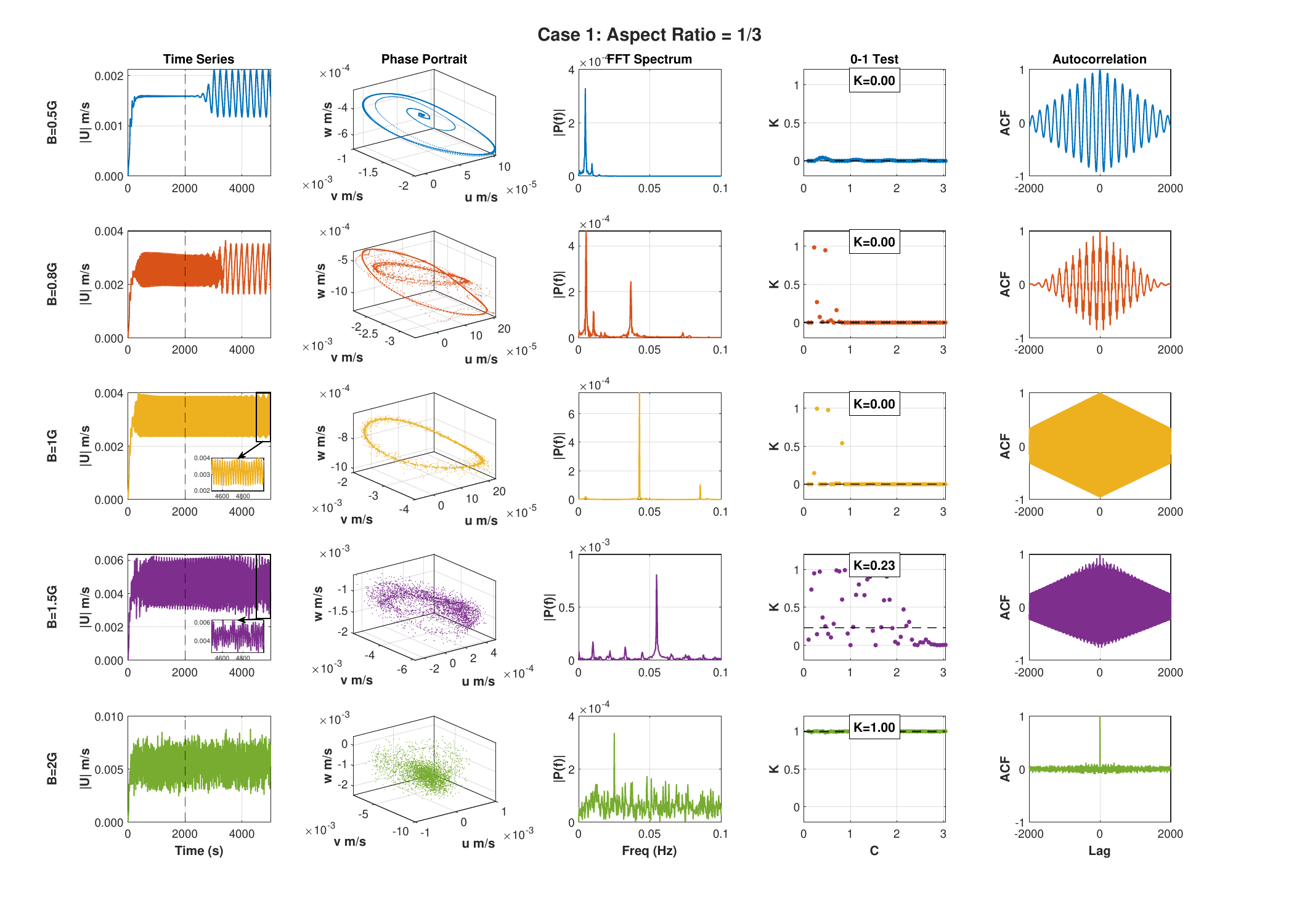}
\caption{Time series, phase portraits, FFT, 0-1 Test for chaos, Autocorrelation function for case1 for aspect ratio 1/3 at different B values}
\label{fig:c1AR13}
\end{figure}    
\end{landscape}

\begin{landscape}
 \begin{figure*}
\centering

\includegraphics[
width=1.5\textwidth,
height=1.5\textheight,
keepaspectratio,trim=1cm 1.5cm 1.5cm 0.5cm,
    clip
]{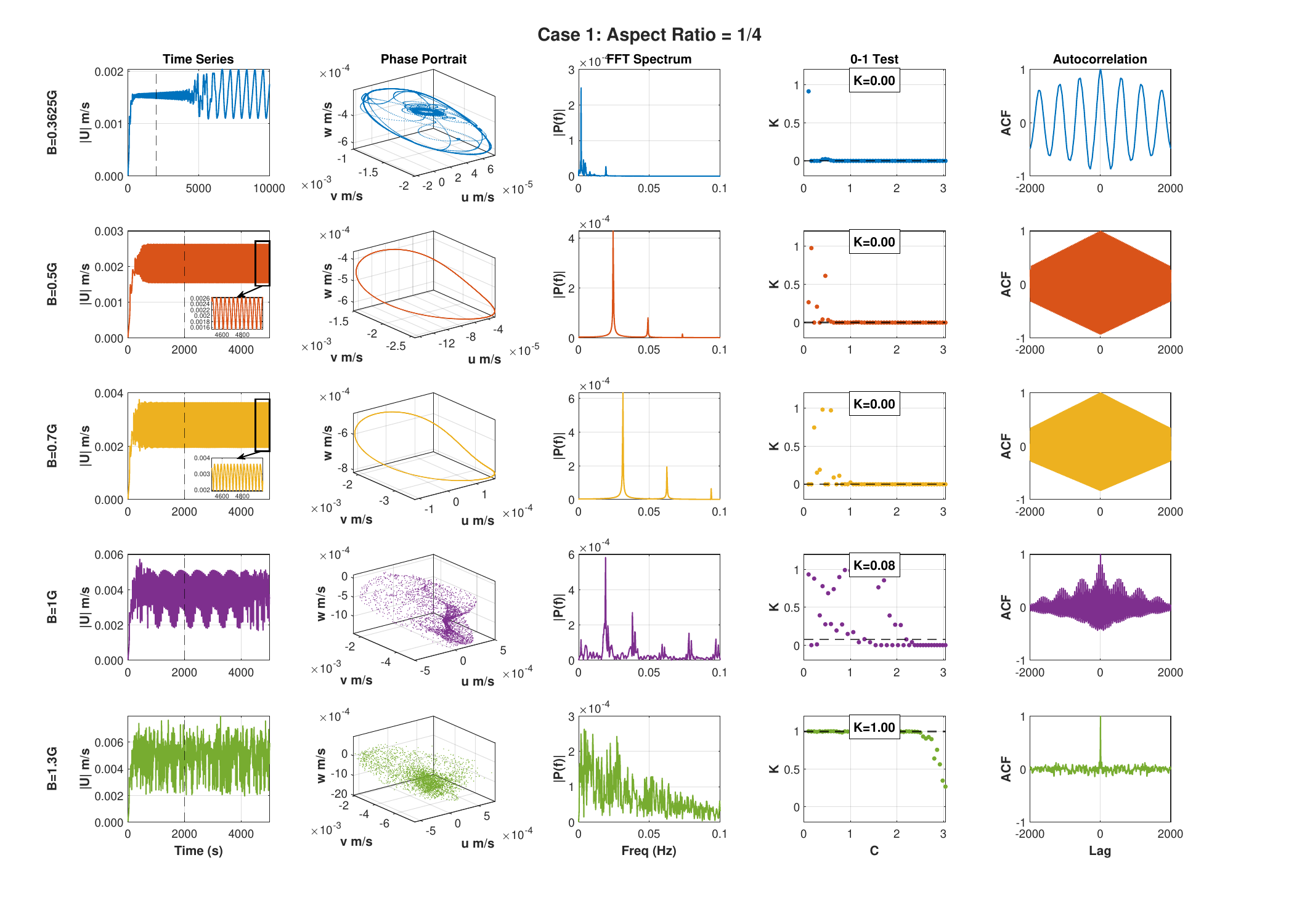}
\caption{
Time series, phase portraits, FFT, 0--1 test for chaos,
and autocorrelation function for case 1 with aspect ratio
$1/4$ at different values of $B$.
}

\label{fig:c1AR14}

\end{figure*}   
\end{landscape}

\begin{landscape}
 \begin{figure}
\centering
\includegraphics[
width=1.5\textwidth,
height=1.5\textheight,
keepaspectratio,trim=1cm 1.5cm 1.5cm 0.5cm,
    clip
]{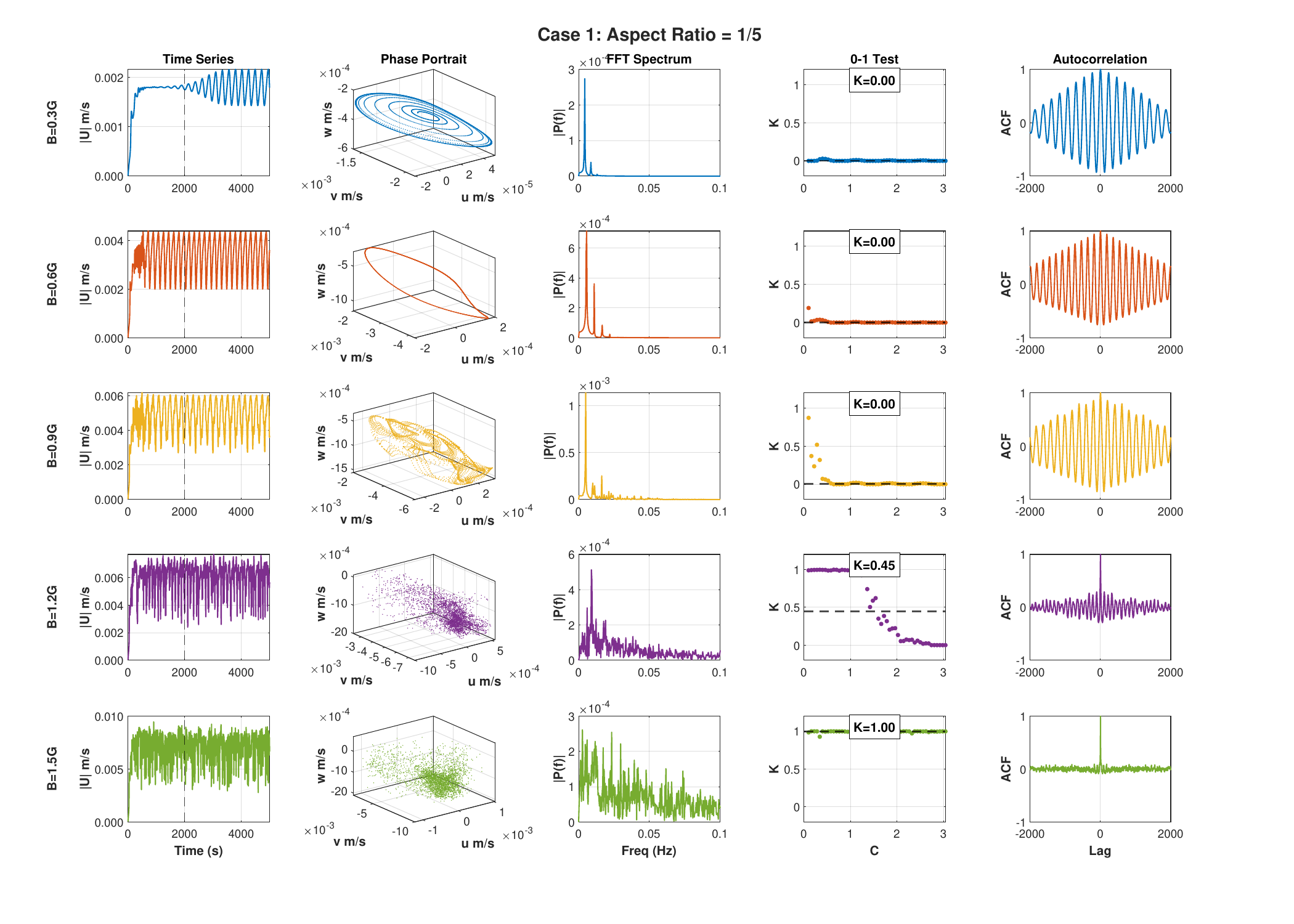}
\caption{Time series, phase portraits, FFT, 0-1 Test for chaos, Autocorrelation function for case1 for aspect ratio 1/5 at different B values}
\label{fig:c1AR15}
\end{figure}   
\end{landscape}

\begin{landscape}
\begin{figure}
\centering
\includegraphics[
width=1.5\textwidth,
height=1.5\textheight,
keepaspectratio,trim=1cm 1.5cm 1.5cm 0.5cm,
    clip
]{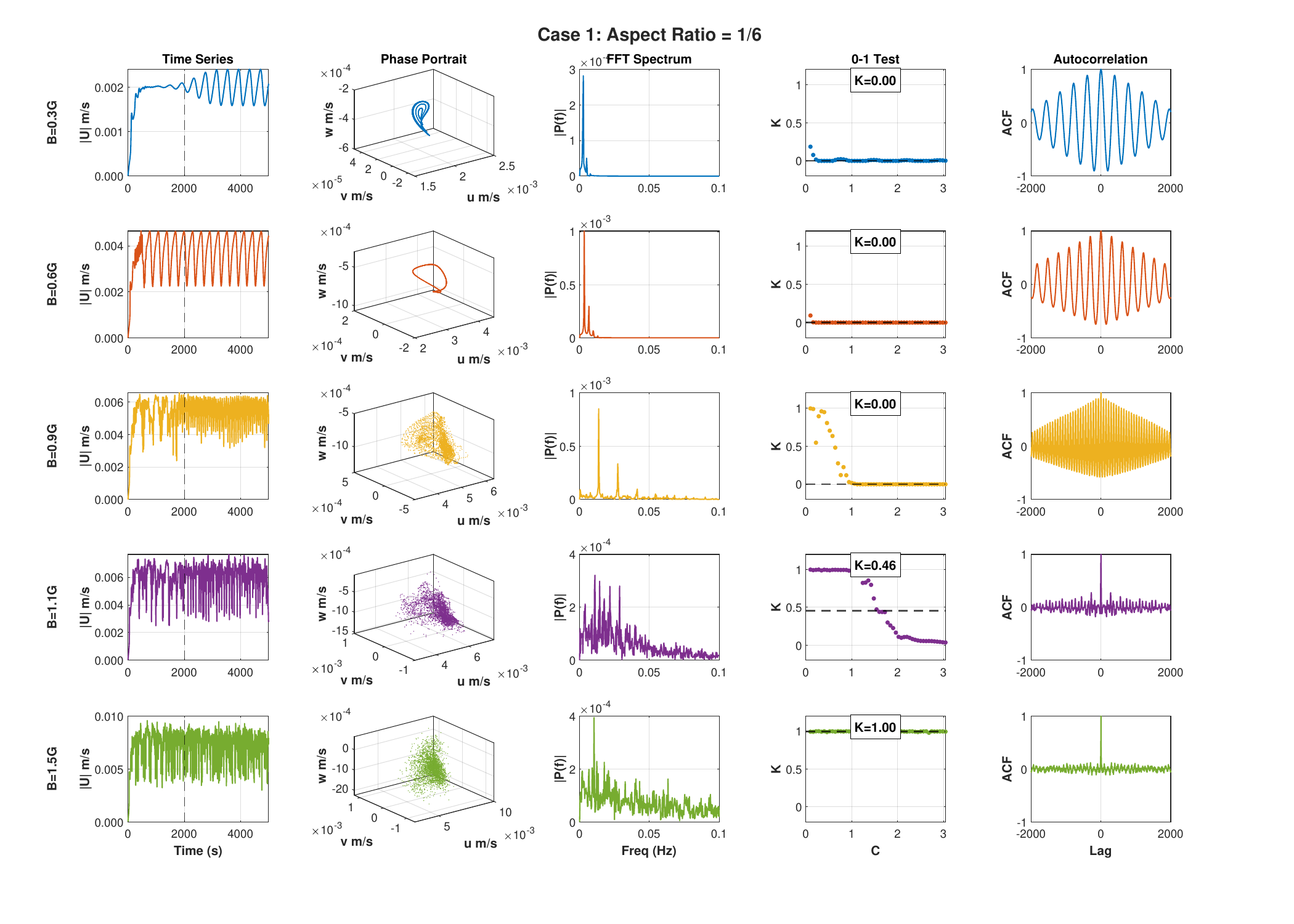}
\caption{Time series, phase portraits, FFT, 0-1 Test for chaos, Autocorrelation function for case1 for aspect ratio 1/6 at different B values}
\label{fig:c1AR16}
\end{figure}    
\end{landscape}

As $B_0$ increases beyond the previous regime (varies for different aspect ratios), the flow  eventually transitions to a chaotic state. As shown in figures ~\ref{fig:c1AR12}--\ref{fig:c1AR16}, a broadband spectrum emerges with multiple interacting frequencies. For values of $B_0$ slightly below this transition, the autocorrelation function (ACF) still exhibits a gradual decay.

At even higher values of $B_0$, the power spectral density (PSD) displays a broad range of frequencies with significant energy, indicating the onset of chaos within this parameter range. With further increase in $B_0$, the flow becomes clearly temporally chaotic at the observation point: the ACF shows rapid de-correlation, and this is corroborated by the 0-1 test for chaos, with the $K$ value approaching unity, confirming chaotic dynamics.

The observed sequence of transitions follows a route consistent with the Ruelle–Takens–Newhouse scenario, with the system evolving from a steady state to periodic oscillations, followed by harmonic enrichment, strange non-chaotic dynamics, and eventually chaos. In this regime, axisymmetry is fully lost for the smaller cell (aspect ratio = 1/2), and the resulting flow is markedly more vigorous compared to the other cases.

\begin{figure}
\centering
\includegraphics[height=0.3\textheight]{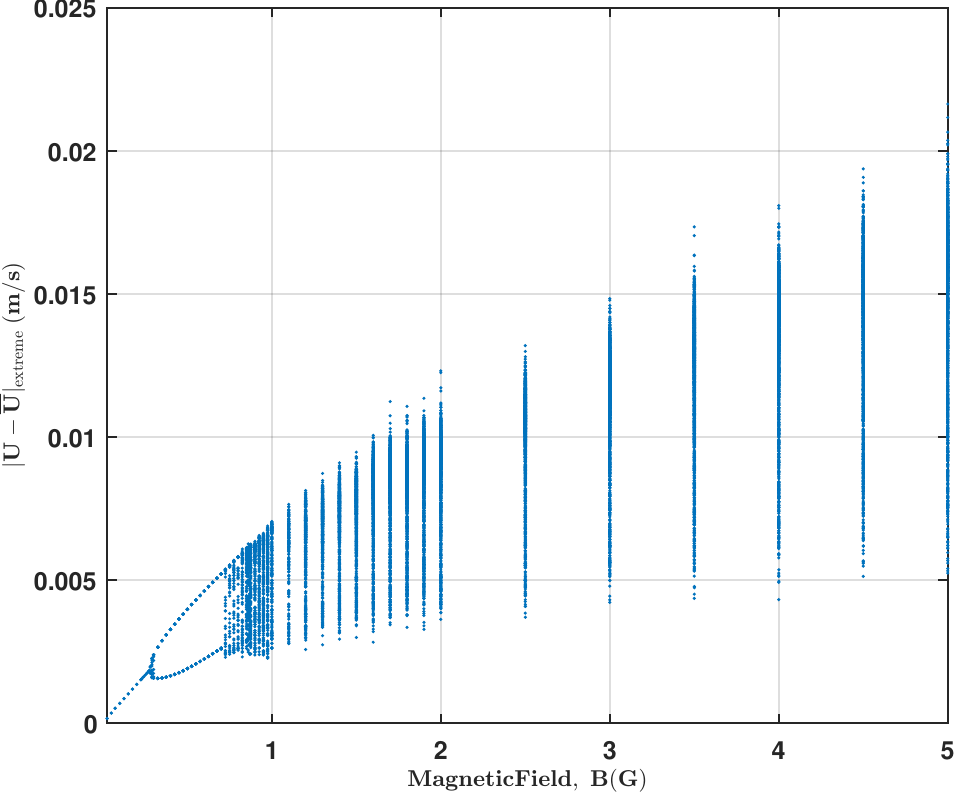}
\captionsetup{width=\textwidth}
\caption{Bifurcation diagram for Aspect ratio = 1/6 for case 1}
\label{fig:c1BD}
\end{figure}

The bifurcation diagram (see figure \ref{fig:c1BD}), plotted for an aspect ratio of 1/6, is constructed using the steady-state values for the steady regime, and the extrema (obtained after subtracting the mean) for all subsequent regimes. The diagram confirms that the transition begins with a supercritical Hopf bifurcation, leading to a quasi-periodic regime characterized by multiple frequency peaks. This is followed by the emergence of strange non-chaotic attractor (SNA) dynamics, and finally by a broadband spectrum with numerous extrema, indicating that the flow has transitioned into a chaotic regime. 
Stability map for this case is shown in figure \ref{fig:stabilitymaps}, which shows that for smaller cells (i.e., aspect ratio = 1/2) the regimes are shifted towards higher ${B_z}$. Even though the current density is higher for this aspect ratio (see table \ref{tab:combinedCR}), a larger ${B_z}$ is needed to reach a given regime as compared to the remaining aspect ratios. For smaller aspect ratios (1/5 and 1/6) the required field values are nearly similar.


\subsection{Modified Chandrasekhar number}\label{section3.2}

After making the $\theta$ momentum equation dimensionless, with the Lorentz-force based on imposed current and magnetic field, and using the velocity scale $U\sim \nu/R$ (expected near the Hopf point) we get 
\begin{equation}
Q_m=\frac{R^{3} J_0 B_0}{\rho \nu^{2}}
\label{eq:modchandeq}
\end{equation}
 which we refer to as the modified Chandrasekhar number, as this quantity is analogous to the Chandrasekhar number,
 \begin{equation}
 Q=Ha^2=\frac{B_0^2 d^2}{\mu_0 \rho \nu \lambda}
 \end{equation}
 where  $\lambda$ is the magnetic diffusivity and $d$ is the characteristic length
 Note that the $Q_m$ is suitable for the cases when both the current density and magnetic field are externally imposed quantities (unlike $Q$ where only magnetic field is externally applied). 
The corresponding values of the modified Chandrasekhar number at the Hopf bifurcation for different aspect ratios are listed in table ~\ref{tab:combinedCR}, and are found to be close. For Cases 1, 2, and 3, the values are of the same order of magnitude, whereas for Case 4 a clear variation is visible, since only the height is changed to vary the aspect ratio, an effect that is not accounted for in the current formulation. Since the kinetic energy imparted by Lorentz force is dissipated at the walls, as H decreases, $\mathbf{J}\times\mathbf{B}$ decreases.

\subsection{Case 2: Effect of $J_0$}
In this case, the geometry is fixed while the $I$ is varied, thereby changing the $J_0$. It is observed that the transition sequence remains unchanged, but the critical magnetic field shifts significantly. For lower currents (smaller current density), a higher background magnetic field is required to trigger instability, as expected (see table ~\ref{tab:combinedCR}, where the Hopf bifurcation occurs at progressively higher values of $B_0$ for lower currents). However, when expressed in terms of the modified Chandrasekhar number, the critical values collapse to a narrow range, confirming that $Q_m$ is the appropriate governing parameter, except when the domain height is changed.

The dynamical evolution of the flow follows the same sequence observed in case 1: steady $\rightarrow$ periodic $\rightarrow$ multi-harmonic $\rightarrow$ SNA $\rightarrow$ chaos. This demonstrates that the transition mechanism is independent of the specific parameter varied. 

\begin{figure*}
\captionsetup{justification=justified,singlelinecheck=false}
\centering

\includegraphics[width=\textwidth]{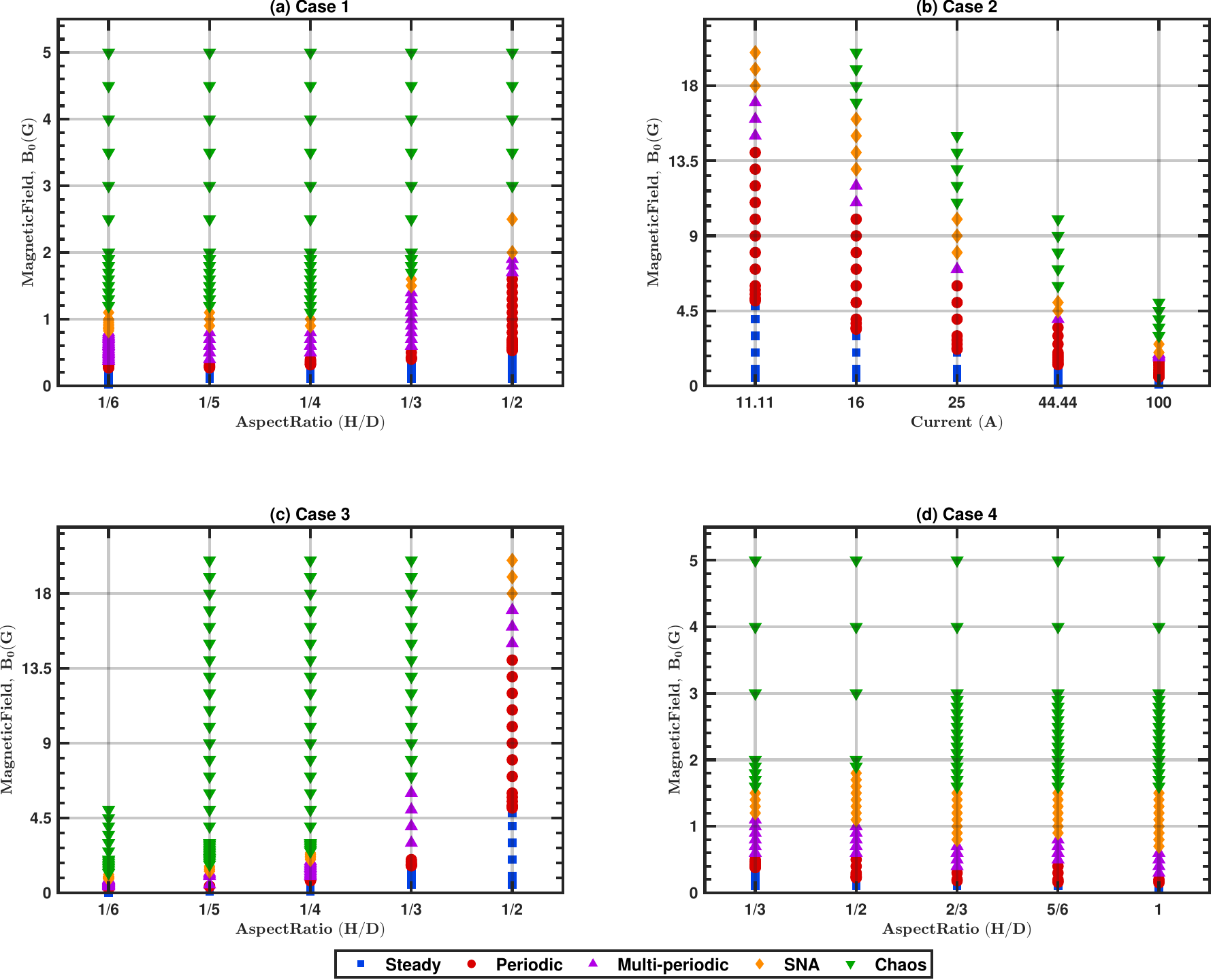}
\captionsetup{width=\textwidth}
\caption{Stability maps for different configurations: 
(a) variation with aspect ratio at fixed current, 
(b) variation with current at fixed geometry, 
(c) variation with aspect ratio at constant current density, 
(d) variation with height at fixed current and radius.}
\label{fig:stabilitymaps}
\end{figure*}

The flow pattern for the different regimes in this case are similar to the corresponding regimes observed for case 1, aspect ratio =1/2. However, the $B_0$ values differ as shown in  table \ref{tab:combinedCR}. The stability map for this case is plotted in figure\ref{fig:stabilitymaps}, case 2. It is clearly observed that, for a given applied vertical magnetic field, the flow regime changes with increasing current density (or current). At lower currents, the flow does not reach the fully turbulent regime even at 20 G, whereas at 100 A, the flow reaches the turbulent regime below 3 G — approximately 15\% of the field strength required in the smallest current case of 11.11 A. This clearly demonstrates that the current rating of the battery is a critical factor in determining the permissible magnetic field exposure during operation; otherwise, the flow may rupture the electrode–electrolyte interface, potentially causing a short circuit within the battery.

\begin{table*}
\centering
\captionsetup{width=\textwidth}
\caption{Critical values of the Hopf bifurcation and Onset of Chaos (OS) for different configurations}

\label{tab:combinedCR}

\begin{tabular}{ccccccccc}
\toprule
\makecell{Radius \\ (m)} &
\makecell{Height \\ (m)} &
\makecell{AR \\ (H/D)} &
\makecell{$I$ \\ (A)} &
\makecell{$B_{\mathrm{Hopf}}$ \\ (G)} &
\makecell{$J$ \\ (A\,m$^{-2}$)} &
\makecell{$Q_m=\frac{R^{3} J_0 B_0}{\rho \nu^{2}}$ \\} &
\makecell{$Re=\frac{\bar{U}_{SS}D}{ \nu}$} &
\makecell{$B_{\mathrm{OC}}$  \\ (G)}  \\
\midrule

\multicolumn{9}{c}{\textbf{Case 1: Effect of Radius, Aspect Ratio and current density}} \\
\midrule
 0.03 & 0.03 & 1/2 & 100 & 0.54--0.55 & 35367.77 & $3.7 \times 10^{5}$ &  1958.05 & 2.5-3\\
 0.045 & 0.03 & 1/3 & 100 & 0.39--0.40 & 15719.01 &$4 \times 10^{5}$ & 2836.15 & 1.6-1.7 \\
 0.06 & 0.03 & 1/4 & 100 & 0.3125--0.315 & 8841.94 & $4.3 \times 10^{5}$ & 3819.52 & 1.1-1.2 \\
 0.075 & 0.03 & 1/5 & 100 & 0.26--0.27 & 5658.84 & $4.4 \times 10^{5}$ & 4310.27 & 1.2-1.3 \\
 0.09 & 0.03 & 1/6 & 100 & 0.26--0.27 & 3929.75 & $5.3 \times 10^{5}$  & 5048.79 & 1.1-1.2\\

\midrule
\multicolumn{9}{c}{\textbf{Case 2: Effect of $J_0$}} \\
\midrule
 0.03 & 0.03 & 1/2 & 11.11 & 5.0--5.1 & 3929.35 & $3.7 \times 10^{5}$ & 1940.23 & -\\
 0.03 & 0.03 & 1/2 & 16 & 3.35--3.45 & 5658.84 & $3.7 \times 10^{5}$ & 1947.37 & 16-17\\
 0.03 & 0.03 & 1/2 & 25 & 2.15--2.25 & 8841.94 & $3.7 \times 10^{5}$ & 1951.38 & 10-11\\
 0.03 & 0.03 & 1/2 & 44.44 & 1.25--1.275 & 15717.43 & $3.6 \times 10^{5}$ &1940.23 & 5-6 \\
 0.03 & 0.03 & 1/2 & 100 & 0.54--0.55 & 35367.76 & $3.7 \times 10^{5}$ & 1958.05 & 2.5-3\\

\midrule
\multicolumn{9}{c}{\textbf{Case 3: Effect of radius for constant $J_0$}} \\
\midrule
 0.03 & 0.03 & 1/2 & 11.11 & 5.0--5.1 & 3929.75 & $3.7 \times 10^{5}$ & 1940.23 & -\\
 0.045 & 0.03 & 1/3 & 25 & 1.6--1.625 & 3929.75 & $4.1 \times 10^{5}$ & 2885.53 & 6-7\\
 0.06 & 0.03 & 1/4 & 44.44 & 0.7--0.8 & 3929.75 & $4.2 \times 10^{5}$  & 3806.82 & 3-4\\
 0.075 & 0.03 & 1/5 & 69.44 & 0.375--0.4 & 3929.75 & $4.5 \times 10^{5}$ & 4353.76 & 2-2.1\\
 0.09 & 0.03 & 1/6 & 100 & 0.26--0.27 & 3929.75 & $5.3 \times 10^{5}$ & 5048.79 & 1.1-1.2\\

\midrule
\multicolumn{9}{c}{\textbf{Case 4: Effect of Height and Aspect Ratio}} \\
\midrule
 0.045 & 0.03 & 1/3 & 100 & 0.375--0.4 & 15719.01 & $4 \times 10^{5}$ & 2836.15 & 1.5-1.6\\
 0.045 & 0.045 & 1/2 & 100 & 0.22--0.225 & 15719.01 & $2.3 \times 10^{5}$ & 1856.75 & 1.8-1.9\\
 0.045 & 0.06 & 2/3 & 100 & 0.175--0.18 & 15719.01 & $1.3 \times 10^{5}$ & 1175.89 & 2-2.1\\
 0.045 & 0.075 & 5/6 & 100 & 0.15--0.16 & 15719.01 & $1 \times 10^{5}$ & 975.36 & 2-2.1\\
 0.045 & 0.09 & 1 & 100 & 0.125--0.15 & 15719.01 & $9.2 \times 10^{4}$ & 888.23 & 1.9-2\\

\bottomrule

\end{tabular}
\end{table*}

\subsection{Case 3: Effect of radius for constant $J_0$}

In this case, both the radius and current are varied simultaneously such that the $J_0=3929.75 A/m^2$ is constant. The results show that the transition behavior remains  identical to the previous cases: the flow undergoes a Hopf bifurcation followed by periodic oscillations, harmonic generation, SNA behavior and eventually  leading to chaos with increase in $B_0$.

Since the current density is fixed, the variation in the critical magnetic field is primarily due to geometric effects. Larger radius leads to reduced viscous damping and hence it requires lower value $B_0$ to trigger instability. However, when expressed in terms of the modified Chandrasekhar number, the critical values again collapse to a similar range (table ~\ref{tab:combinedCR}).

The flow pattern remains similar to that in the case 1, since aspect ratios is fixed. Interestingly, however,  the effect of $J$ is found to be more significant than that of current, $I$. In case 1, the onset of chaos occurs within a relatively narrow range of 1.2–3 G across the different aspect ratios, whereas in the present case, this range is considerably wider: for the aspect ratio 1/2, SNA is still observed even at 20 G, while for aspect ratio 1/6, chaos sets in as early as 1.2 G. The Hopf point is likewise shifted significantly — for aspect ratio 1/2, it occurs between 5–5.1 G in this case, compared to 0.54–0.55 G in case 1 for the same aspect ratio. This inverse relationship between $B_z$ and $I$ id further confirmed by the modified Chandrashekar number, given in Eq.\ref{eq:modchandeq}.

\subsection{Case 4: Effect of Height and Aspect Ratio}

In this case, the height is varied while keeping the radius and current constant. This modifies the aspect ratio without altering the $J_0$.

The transition sequence remains consistent with the previous cases. The flow is initially steady and undergoes a Hopf bifurcation as $B_0$ increases. This is followed by periodic oscillations, harmonic enrichment, SNA behavior and eventual transition to chaos.

Increasing the height alters the vertical confinement of the flow, which influences the viscous dissipation as well as the spatial flow structure. As a result, the critical magnetic field varies with aspect ratio, as shown in table ~\ref{tab:combinedCR}. However, the corresponding values of the modified Chandrasekhar number is not as close as the remaining cases.

\begin{figure*}
\captionsetup{justification=justified,singlelinecheck=false}
\centering
\includegraphics[width=\textwidth]{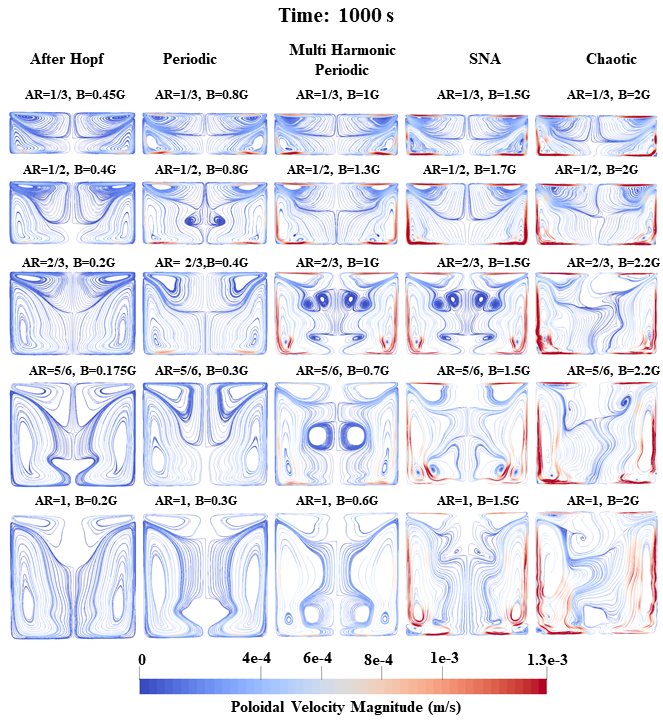}
\captionsetup{width=\textwidth}

\caption{Instantaneous Poloidal Velocity magnitude for Aspect Ratio 1/3, 1/2, 2/3, 5/6,1 with applied current 100A at 4000 Seconds at different Background Magnetic fields for case 4 }
\label{fig:UPol_C4}
\end{figure*}

From figure ~\ref{fig:UPol_C4}, it is clear that a simple, clearly separated primary–secondary flow structure observed across all regimes only for aspect ratio 1/3. For the remaining aspect ratios, the flow structure is slightly different, a distorted flow pattern is observed, along with the formation of a ``recirculation bubble'' similar to that seen in case 1, (aspect ratio 1/2). For aspect ratio 2/3, during the multi-harmonic and SNA regimes; a similar behaviour is observed for the remaining two larger aspect ratios as well. For all aspect ratios greater than or equal to 1/2, axisymmetry is destroyed once the flow reaches the turbulent regime.

The stability map for this case is shown in figure ~\ref{fig:stabilitymaps}, case 4. Here it can be observed that although the Hopf point differs significantly across aspect ratios (see table ~\ref{tab:combinedCR}), the onset of the chaotic regime occurs at a nearly consistent field strength for all aspect ratios, close to 1.6 G. This suggests that, when designing larger cells at constant current, it is preferable to increase the height of the domain rather than its radius. This however, may trigger Tayler instability \cite{STEFANI20112982, Herreman2015}.

\subsection{Unified dynamical behaviour}

Across all four cases, the flow exhibits a consistent sequence of dynamical transitions:
\begin{equation}
\text{steady} \;\rightarrow\; \text{periodic} \;\rightarrow\; \text{multi-harmonic} \;\rightarrow\; \text{SNA} \;\rightarrow\; \text{chaos}.
\end{equation}

This sequence is broadly consistent with a Ruelle--Takens--Newhouse-type route to chaos, with the presence of a strange non-chaotic attractor as an intermediate regime.



\subsection{Scaling laws }\label{section3.7}

All simulations satisfy the criterion proposed by \cite{Davidson1999} for swirl-dominated motion, the resulting flow is primarily governed by the azimuthal circulation rather than the poloidal electro-vortex flow. 

For swirling flows, several scaling laws have been proposed depending on the dominant force balance.
In the inductionless limit, different balances between Lorentz, viscous and inertial forces lead to different power-law relationships. A viscous-Lorentz balance predicts

\begin{equation}
U \sim \frac{JBR^{2}}{\rho\nu}\quad or\quad Re \sim Q_m
\end{equation}

while an Lorentz - inertial balance gives

\begin{equation}
U \sim \left(\frac{JBR}{\rho}\right)^{1/2} \quad or\quad Re \sim Q_m^{1/2}
\end{equation}

To capture the exchange of momentum and energy occur between the boundary layer and the bulk regions \cite{Herreman2021} has proposed 2/3 scaling law. This depends on a three way balance in the Z-vorticity equation.

\begin{figure*}
\captionsetup{justification=justified,singlelinecheck=false}
\centering
\includegraphics[width=1\textwidth]{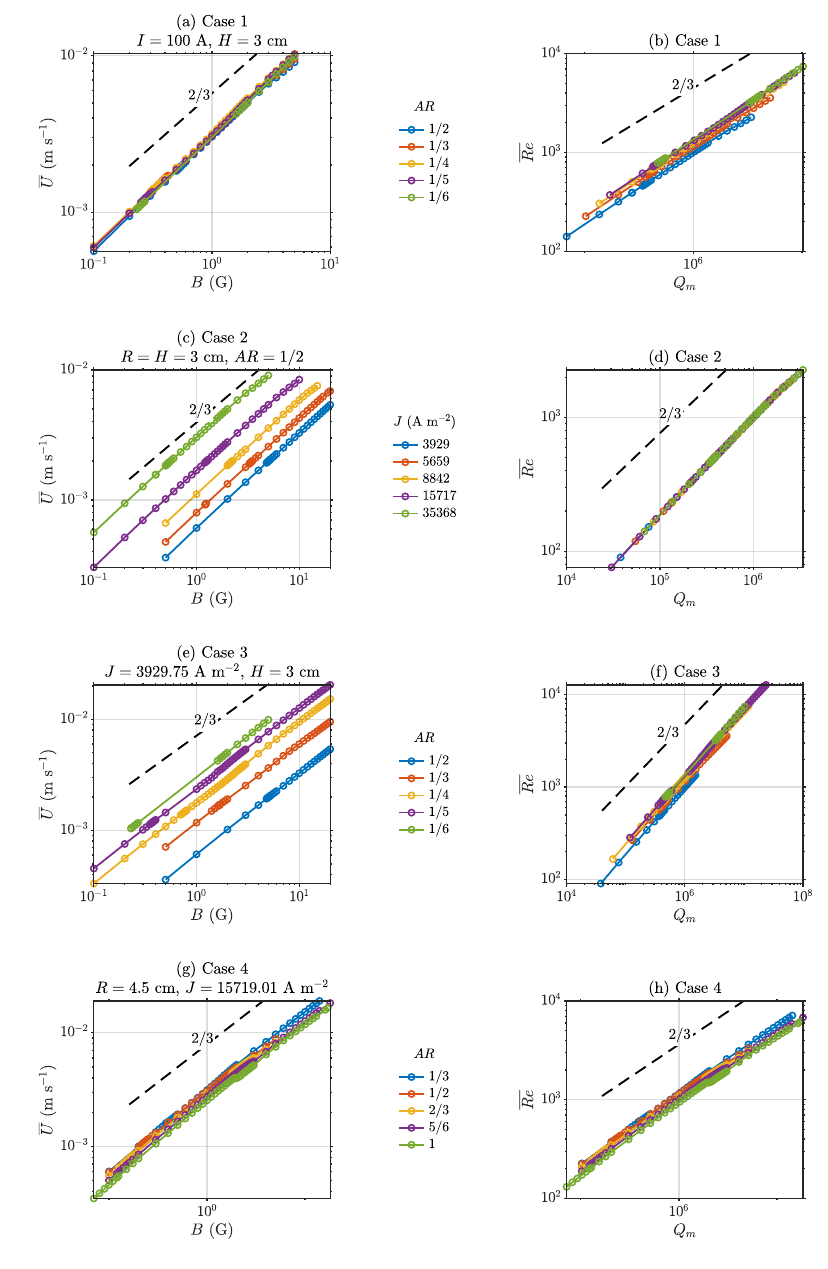}
\captionsetup{width=\textwidth}
\caption{Time and space averaged velocity as a function of B  and $\bar Re$ vs $Q_m$ for all the four cases}
\label{fig:Scaling}
\end{figure*}

\begin{equation}
U \sim
\left(\frac{JB}{\rho}\right)^{2/3}
\frac{R}{\nu^{1/3}}\quad or\quad Re \sim Q_m^{2/3}
\end{equation}

Earlier, \cite{Davidson1992} had proposed 2/3 scaling based on global torque balance. Recently, \cite{benard2022} have shown that 2/3 and 1/2 work over different regimes.

To assess whether these scaling relations remain valid over the range of geometries and operating conditions considered in the present study, the time- and space-averaged velocity was analysed as a function of both the applied magnetic field and the current density.

\begin{figure*}
\captionsetup{justification=justified,singlelinecheck=false,width=\textwidth}
\centering
\includegraphics[width=\textwidth,
height=0.3\textheight, keepaspectratio]{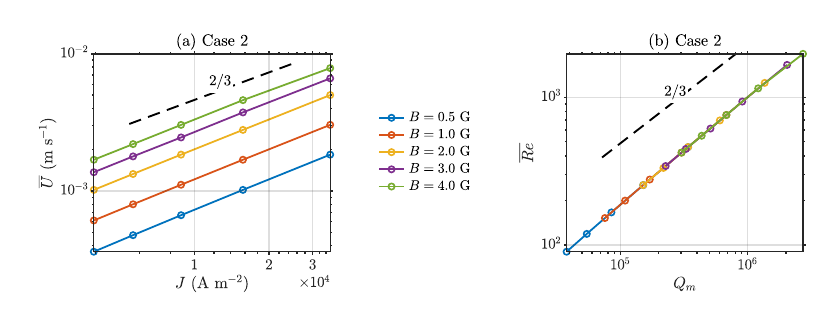}
\captionsetup{width=\textwidth}
\caption{Time and space averaged velocity as a function of J and $\bar Re$ vs $Q_m$ for the Second case }
\label{fig:ScalingC2J}
\end{figure*}

Figures ~\ref{fig:Scaling} and ~\ref{fig:ScalingC2J} present the variation of the mean velocity with the applied magnetic field and current current densities respectively along with the variation of $\bar Re$ with $Q_m$ for all the cases. In every configuration, the numerical results exhibit the expected 2/3 power-law dependence,

\begin{equation}
\bar{U}\propto B^{\,n}, \quad and \quad \bar{U}\propto J^{n} \implies \bar{Re}\propto{Q_m}^n
\end{equation}

From Figures ~\ref{fig:Scaling} (a,c,e,g) and ~\ref{fig:ScalingC2J} (a) show a clear separation between different cases, indicating a strong dependence on the individual geometric and forcing parameters, the corresponding $\bar Re$ vs $Q_m$ plots exhibit a collapse onto a common line. This suggests that $Q_m$, provides an appropriate characterization of the flow across the different cases. 
where the fitted exponent varies only slightly with geometry and operating conditions. The corresponding exponents and coefficients obtained from the least-squares fits are summarized in table ~\ref{tab:Scaling}. For all the cases the exponent lies between approximately $0.69$ and $0.74$ for all cases, very close to the 2/3rd scaling expected in this regime. This suggests that the flow is governed by a combination of Lorentz, inertial and viscous effects, while also being influenced by geometry, for example the location of the current collectors. Variations in aspect ratio alter the relative importance of these competing mechanisms, resulting in small but systematic changes in the scaling exponents, while preserving the overall power-law dependence of the mean velocity.

\begin{table*}
\centering
\captionsetup{width=\textwidth}
\caption{Power-law scaling relations for the mean velocity. Current-density scaling follows
$\bar{u}=CJ^{n}$, while magnetic-field scaling follows
$\bar{u}=aB^{n}$.}
\label{tab:Scaling}

\begin{tabular}{ccccccc}
\toprule
Radius &
Height &
AR &
$I$ &
$J$ &

Exponent ($n$) &
$R^2$\\
(m) & (m) & (H/D) & (A) & (A\,m$^{-2}$) &  &\\

\midrule
\multicolumn{7}{c}{\textbf{Magnetic field scaling: $\bar{u}=aB^{n}$}}\\
\midrule

\multicolumn{7}{c}{\textbf{Case 1: Effect of Radius, Aspect Ratio and current density}} \\
\midrule
0.03 & 0.03 & 1/2 & 100 & 35367.77  & 0.7125 & 0.99787\\
0.045& 0.03 & 1/3 & 100 & 15719.01  & 0.7044 & 0.99950\\
0.06 & 0.03 & 1/4 & 100 & 8841.94  & 0.7268 & 0.99983\\
0.075& 0.03 & 1/5 & 100 & 5658.84   & 0.7272 & 0.99942\\
0.09 & 0.03 & 1/6 & 100 & 3929.75   & 0.7303 & 0.99987\\

\midrule
\multicolumn{7}{c}{\textbf{Case 2: Effect of $J_0$}} \\
\midrule
0.03&0.03&1/2&11.11&3929.35  &0.7296&0.99989\\
0.03&0.03&1/2&16&5658.84 &0.7241&0.99973\\
0.03&0.03&1/2&25&8841.94 &0.7165&0.99903\\
0.03&0.03&1/2&44.44&15717.43 &0.7213&0.99799\\
0.03&0.03&1/2&100&35367.76 &0.7125&0.99787\\

\midrule
\multicolumn{7}{c}{\textbf{Case 3: Effect of radius for constant $J_0$}} \\
\midrule
0.03&0.03&1/2&11.11&3929.75 &0.7296&0.99989\\
0.045&0.03&1/3&25&3929.75 &0.7017&0.99969\\
0.06&0.03&1/4&44.44&3929.75 &0.7230&0.99970\\
0.075&0.03&1/5&69.44&3929.75 &0.7202&0.99981\\
0.09&0.03&1/6&100&3929.75 &0.7303&0.99987\\

\midrule
\multicolumn{7}{c}{\textbf{Case 4: Effect of Height and Aspect Ratio}} \\
\midrule
0.045&0.03&1/3&100&15719.01&0.6964&0.99896\\
0.045&0.045&1/2&100&15719.01&0.6920&0.99637\\
0.045&0.06&2/3&100&15719.01&0.7001&0.99530\\
0.045&0.075&5/6&100&15719.01&0.6935&0.99521\\
0.045&0.09&1&100&15719.01&0.7012&0.99539\\

\midrule
\multicolumn{7}{c}{\textbf{Current density scaling: $\bar{u}=CJ^{n}$}}\\
\midrule
Radius &
Height &
AR &

B &
Exponent ($n$) &
$R^2$\\
(m) & (m) & (H/D) & (G) & &\\

\midrule
\multicolumn{6}{c}{\textbf{Case 2: Effect of $J_0$}} \\
\midrule
0.03 & 0.03 & 1/2  &
0.5 &
0.7415 &
0.99991\\

0.03 & 0.03 & 1/2 &
1 &
0.7304 &
0.99996\\

0.03 & 0.03 & 1/2  &
2 &
0.7243 &
0.99999\\

0.03 & 0.03 & 1/2 &
3 &
0.7177 &
0.99992\\

0.03 & 0.03 & 1/2 &
4 &
0.7023 &
0.99922\\

0.03 & 0.03 & 1/2&
5&
0.6948 &
0.99855\\
\bottomrule
\end{tabular}
\end{table*}

\section{Conclusions}\label{section4}

The transition to unsteady and chaotic flow in an electromagnetically driven liquid metal system has been investigated using extensive three-dimensional numerical simulations. By systematically varying the background magnetic field across a range of aspect ratios and current density (give range), the sequence of dynamical transitions leading to chaos has been characterised.

Across all cases, the flow exhibits a consistent progression from a steady state to periodic oscillations via a Hopf bifurcation. With increasing magnetic field strength, the dynamics evolve through a multi-harmonic periodic regime, followed by the emergence of strange non-chaotic attractors, and ultimately transitions to fully developed chaos. The identification of strange non-chaotic attractors as an intermediate regime highlights the presence of complex dynamics that are not captured by conventional classifications based solely on periodic or chaotic behaviour.

The critical parameter for the onset of instability, when expressed in terms of the dimensionless number,
\begin{equation}
Q_m = \frac{R^{3} J_0 B_0}{\rho \nu^{2}},
\end{equation}
which we term as the modified Chandrasekhar number, collapses to a narrow range across all cases (except when the domain height is changed). This demonstrates that the transition is governed by a particular balance between Lorentz, and viscous forces. The onset of chaos occurs at $1.1-1.2 G$ for a aspect ratio $1/6$ and a current of 100A, as suggested earlier in \cite{davidson2022swirling}, but this is shifted to $16-17 G$ for a lower current of 16A and aspect ratio $1/2$. At a lower current, there is no chaotic behaviour seen even at $20 G$. 

The observed transition pathway is broadly consistent with a Ruelle--Takens--Newhouse-type route to chaos, with modifications arising from the electromagnetic nature of the forcing. In particular, the presence of harmonic enrichment prior to the onset of strange non-chaotic dynamics distinguishes this system from classical quasi-periodic routes to chaos.

Overall, the present study demonstrates that electromagnetically driven flows in liquid metal systems exhibit a robust and universal route to chaos, governed by the complex interplay between the kinetic energy imparted by the Lorentz forcing and viscous dissipation. 
In real LMBs, however, multiphysics effects due to interfacial motion (\cite{herreman2019}) and solutal stratification (\cite{Herreman2021}) are expected to be important. The effect of these factors on transition to turbulence will be taken up in a future study.

\backsection[Acknowledgements]{AR wishes to thank Prof. Davidson for the inspiration to work on this problem. The authors gratefully acknowledge Prof. Venkatasailanathan Ramadesigan for providing access to the computational resources. The authors also thank Ashutosh Narayan Singh for the helpful discussions on methods for nonlinear dynamical analysis. The authors also acknowledge National Supercomputing Mission (NSM) for providing computing
resources of ‘PARAM RUDRA’ at G02, Ground Floor, IIT Bombay, Powai, Mumbai, 400076,
which is implemented by C-DAC and supported by the Ministry of Electronics and
Information Technology (MeitY) and Department of Science and Technology (DST),
Government of India.}

\backsection[Declaration of interests]{The authors report no conflict of interest.}

\appendix

\section{}\label{appA}

Consider the  equation, \eqref{eq:harmonic} in cylindrical polar coordinates assuming axisymmetry
For which, using the variable separation with $\phi(r,z)=R(r)Z(z)$, we get

\begin{equation}
    rR^{\prime\prime}+R^{\prime}+\lambda^2rR=0
    \label{Reqn}
\end{equation}
\begin{equation}
    Z^{\prime\prime}-\lambda^2Z=0
    \label{Zeqn}
\end{equation}

Here $\lambda^2$ is a constant.

Now the potential can be expressed as follows
\begin{equation}
    \phi(r,z)=R(r)Z(z)=C_1J_0(\lambda r)\times\left(D_1cosh(\lambda z)+D_2 sinh(\lambda z)\right)
    \label{phieq}
\end{equation}

Where $J_0, J_1$ are Bessel functions of first and second kind.
At the bottom boundary, z=0, the surface is insulated so
\begin{equation}
    \left.J_z\right|_{z=0}=-\sigma\left.\frac{\partial \phi}{\partial z}\right|_{z=0}=-\sigma \left(C_1 J_0(\lambda r)\right)(D_2 ) =0
    \label{Jzatz=0}
\end{equation}
For the above equation \eqref{Jzatz=0} to be true for every r, either $C_1$ or $D_2$ should be zero, however $C_1$ cannot be zero then the potential will be independent of r, so $D_2$ should be zero. Now the formulation for potential will be as follows
\begin{equation}
 \phi(r,z)=CJ_0(\lambda r)cosh(\lambda z) 
 \label{phi}
\end{equation}

From \eqref{eq:stokesohm} and \eqref{phi}, the Stream function for current density can be calculated as follows
\begin{equation}
    J_z=-\sigma \frac{\partial \phi}{\partial z}=-\sigma C\lambda J_0(\lambda r)sinh(\lambda z)=\frac{1}{r}\frac{\partial \Psi}{\partial r}
    \label{combJ}
\end{equation}
Integrating \eqref{combJ} with respect to r and using the identity $\int xJ_0(x)dx=xJ_1(x)$ can give $\Psi$
\begin{equation}
    \Psi=-\sigma Csinh(\lambda z)\frac{1}{\lambda}[(\lambda r)J_1(\lambda r)]=-\sigma CrJ_1(\lambda r) sinh(\lambda z)
    \label{Phi1}
\end{equation}
The boundary condition at the outer wall (r=$R_0$) requires no current flow through domain radially, which means $\Psi$ should be zero on the wall.
\begin{equation}
    \Psi(R_0,z)=-\sigma CR_0J_1(\lambda R_0) sinh(\lambda z)=0
    \label{Phi2}
\end{equation}

Now the general solution for potential will be 
\begin{equation}
    \phi(r,z)=\sum_{n=1}^{\infty}A_nJ_0(\frac{\delta_n r}{R_0}) cosh(\frac{\delta_n z}{R_0})
    \label{Phi3}
\end{equation}
\subsection*{Determination of Coefficients via Orthogonality}

The coefficients $A_n$ are determined by the potential distribution prescribed at the top boundary ($z=D$). Let the prescribed potential profile be $f(r)$.
\begin{equation}
    \phi(r, D) = \sum_{n=1}^{\infty} A_n \cosh\left( \frac{\delta_n D}{R_0} \right) J_0\left( \frac{\delta_n r}{R_0} \right) = V f(r)
\end{equation}
This is a Fourier-Bessel series expansion. To isolate $A_n$, we utilize the orthogonality property of Bessel functions. For functions $J_0(\lambda_n r)$ where $\lambda_n$ are roots of $J_1(\lambda R_0)=0$, the orthogonality condition is:
\begin{equation}
    \int_0^{R_0} r J_0\left( \frac{\delta_n r}{R_0} \right) J_0\left( \frac{\delta_m r}{R_0} \right) dr = \frac{R_0^2}{2} [J_0(\delta_n)]^2 \delta_{nm}
\end{equation}
Multiplying both sides of the summation by $r J_0(\frac{\delta_m r}{R_0})$ and integrating from $0$ to $R_0$:
\begin{equation}
    A_n \cosh\left( \frac{\delta_n D}{R_0} \right) \frac{R_0^2}{2} J_0^2(\delta_n) = V \int_0^{R_0} r f(r) J_0\left( \frac{\delta_n r}{R_0} \right) dr
\end{equation}
Solving for $A_n$:
\begin{equation}
    A_n = \frac{2 V}{R_0^2 J_0^2(\delta_n) \cosh(\frac{\delta_n D}{R_0})} \int_0^{R_0} r f(r) J_0\left( \frac{\delta_n r}{R_0} \right) dr
\end{equation}
The integral term depends on the specific shape of $f(r)$. For the profile defined in Section 3.2 (uniform up to $R-T$, then linear decay to $R$), the integration yields the geometric factor $\tilde{J}_1(n)$ defined in Eq. (6) of the \cite{davidson2022swirling}:
\begin{equation}
    \tilde{J}_1(n) = \frac{1}{RT} \int_{R-T}^{R} r J_1\left( \frac{\delta_n r}{R_0} \right) dr
    \label{Jtilda}
\end{equation}
Substituting this back, we obtain the explicit expression for the coefficients:
\begin{equation}
    A_n = \frac{2 V R \tilde{J}_1(n)}{R_0 \delta_n J_0^2(\delta_n) \cosh(\delta_n D / R_0)}
    \label{An}
\end{equation}
This completes the derivation of the electrostatic potential field.



The azimuthal component of Lorentz force is $F_\theta  = {J_r B_0}/{\rho} $, with vertical magnetic field and poloidal current density ($\mathbf{J} = J_r \hat{e}_r + J_z \hat{e}_z$) :

We now require the explicit expression for $J_r$. 
\begin{equation}
    J_r(r,z) = -\frac{1}{r} \frac{\partial \Psi}{\partial z}=\sigma \sum_{n=1}^{\infty} A_n \frac{\delta_n}{R_0} J_1\left( \frac{\delta_n r}{R_0} \right) \cosh\left( \frac{\delta_n z}{R_0} \right)
\end{equation}
Now, substituting $J_r$ into the Lorentz force and replacing $A_n$ using \ref{An} with its explicit form involving $V$:
\begin{equation}
    F_\theta(r,z) = \frac{2 \sigma V B_0}{\rho} \frac{R}{R_0^2} \sum_{n=1}^{\infty} \frac{\tilde{J}_1(n) J_1(\delta_n r / R_0)}{J_0^2(\delta_n)} \frac{\cosh(\delta_n z / R_0)}{\cosh(\delta_n D / R_0)}
\end{equation}

\section{}\label{appB}

\subsection{The 0--1 Test for Chaos}\label{appB.1}

The 0--1 test for chaos was originally proposed by Gottwald \& Melbourne
\cite{gottwald2004} and subsequently refined in \cite{gottwald2009}. Unlike
methods that require phase-space reconstruction, it operates directly on a
scalar time series and returns a single indicator $K \in [0,1]$, where $K
\approx 0$ indicates regular (periodic or quasi-periodic) dynamics and $K
\approx 1$ indicates chaotic dynamics.

\subsubsection{Pre-processing}

Before applying the test, the raw three-component velocity signal
$(u, v, w)$ is processed as follows. The velocity magnitude
\begin{equation}
    |\mathbf{U}|(t_i) = \sqrt{u_i^2 + v_i^2 + w_i^2}
\end{equation}
is computed at each time step. The first $N_\mathrm{tr} = 2000$ samples are
discarded to remove the initial transient, yielding the steady-state signal
$\phi_j = |\mathbf{U}|_{N_\mathrm{tr}+j}$. The signal is then mean-subtracted
and standardised,
\begin{equation}
    \phi_j^* = \frac{\phi_j - \langle\phi\rangle}{\sigma_\phi},
\end{equation}
where $\langle\phi\rangle$ and $\sigma_\phi$ denote the sample mean and
standard deviation respectively. This normalisation ensures that the subsequent
displacement variables remain bounded regardless of the physical amplitude of
the signal, and removes any constant offset that would otherwise produce a
spurious linear growth in the translation variables.

\subsubsection{Translation Variables}

For a given parameter $c \in (0, \pi)$, the two-dimensional translation
variables are defined by the cumulative sums
\begin{equation}
    p(n,c) = \sum_{j=1}^{n} \phi_j^* \cos(jc), \qquad
    q(n,c) = \sum_{j=1}^{n} \phi_j^* \sin(jc),
\label{eq:pq}
\end{equation}
for $n = 1, 2, \ldots, N$, where $N$ is the length of the trimmed signal.
The key insight of the method is that the asymptotic behaviour of the
$(p,q)$ trajectory distinguishes between dynamical regimes: for regular
dynamics the trajectory remains bounded, while for chaotic dynamics it
executes a Brownian-like random walk and grows without bound.

\subsubsection{Modified Mean Square Displacement}

To quantify this growth, the mean square displacement (MSD) is computed as
\begin{equation}
    M(n,c) = \frac{1}{N-n}\sum_{j=1}^{N-n}
              \Bigl[\bigl(p(j+n)-p(j)\bigr)^2 +
                    \bigl(q(j+n)-q(j)\bigr)^2\Bigr],
\label{eq:Mnc}
\end{equation}
evaluated over a lag window $n = 1, 2, \ldots, \lfloor N/10 \rfloor$.
For regular dynamics $M(n,c)$ saturates to a constant, whereas for
chaotic dynamics it grows linearly with $n$.

\subsubsection{The Growth Rate $K$}

The growth rate associated with a particular value of $c$ is estimated by
the Pearson correlation coefficient between the integer lag vector
$\mathbf{n} = (1, 2, \ldots, n_\mathrm{max})^\top$ and the corresponding
displacement vector $\mathbf{M}(c)$:
\begin{equation}
    K(c) = \mathrm{corr}\bigl(\mathbf{n},\, \mathbf{M}(c)\bigr).
\label{eq:Kc}
\end{equation}
$K(c) \approx 1$ when $M$ grows linearly (chaotic case), and $K(c) \approx 0$
when $M$ is flat (regular case). Negative values can arise from resonances
between $c$ and the dominant frequency of a quasi-periodic signal; these are
set to zero to avoid spurious results.

To suppress such resonance artifacts, the test is repeated for a grid of
50 values of $c$ uniformly distributed over $(0.1,\, \pi - 0.1)$, and the
final chaos indicator is taken as the median,
\begin{equation}
    K = \mathrm{median}_{c}\bigl\{K(c)\bigr\}.
\label{eq:Kmed}
\end{equation}
The median is preferred over the mean because a small number of resonant
$c$ values can drive $K(c)$ strongly negative, which would distort an
arithmetic average but has negligible effect on the median
\cite{gottwald2009}. In the present study the threshold $K \geq 0.8$ is
used to classify a signal as chaotic, consistent with the criterion
adopted in similar fluid-dynamics applications
\cite{falconer2007}.

\subsection{Correlation Dimension} \label{appB.2}

The correlation dimension $D_2$ is a measure of the fractal dimension of
the attractor embedded in phase space. It is estimated using the algorithm
of Grassberger \& Procaccia \cite{grassberger1983}, which proceeds in
three stages: (i) reconstruction of the attractor from the scalar time
series via time-delay embedding, (ii) computation of the correlation
integral, and (iii) extraction of $D_2$ from the scaling exponent of that
integral.

\subsubsection{Time-Delay Embedding}

Takens' embedding theorem \cite{takens1981} guarantees that the topology
of the original attractor can be faithfully recovered from a scalar
observable $\phi$ provided the embedding dimension $m$ and time lag
$\tau$ are chosen appropriately. The delay-coordinate vectors
\begin{equation}
    \mathbf{y}_i = \bigl(\phi_i,\, \phi_{i+\tau},\, \ldots,\,
                         \phi_{i+(m-1)\tau}\bigr)^\top, \quad
    i = 1, \ldots, N - (m-1)\tau,
\label{eq:embed}
\end{equation}
define a cloud of points in $\mathbb{R}^m$ that traces out the
reconstructed attractor.

\textbf{Lag selection via mutual information.}
The time lag $\tau$ is chosen as the first local minimum of the
auto-mutual information function
\begin{equation}
    I(\tau) = \sum_{r,s} P(r,s;\tau)
              \ln\!\left[\frac{P(r,s;\tau)}{P_1(r)\,P_1(s)}\right],
\label{eq:MI}
\end{equation}
where $P(r,s;\tau)$ is the joint probability that $\phi$ falls in bin $r$
at time $t$ and bin $s$ at time $t+\tau$, and $P_1(r)$, $P_1(s)$ are the
corresponding marginal probabilities, estimated from equal-width
histogram bins. The first minimum of $I(\tau)$ identifies the lag at which
successive delay coordinates carry the least mutual information---i.e.\ they
are as statistically independent as possible---which maximally ``unfolds''
the attractor and minimises redundancy in the reconstructed coordinates.
If no minimum is found within the search range $\tau \in [1, 50]$, a default
value of $\tau = 10$ is used.

The minimum sufficient embedding dimension is determined by the method of
false nearest neighbours (FNN) \cite{kennel1992}. For each candidate
dimension $m$, the FNN fraction is defined as
\begin{equation}
    \mathrm{FNN}(m) = \frac{1}{N_\mathrm{test}}
    \sum_{i}\mathbf{1}\!\left[
        \frac{|\phi_{i+m\tau}-\phi_{\hat\imath+m\tau}|}{d_m(i,\hat\imath)}
        > R_\mathrm{tol}
    \right],
\label{eq:FNN}
\end{equation}
where $\hat\imath$ is the index of the nearest neighbour of $\mathbf{y}_i$
in $m$ dimensions, $d_m(i,\hat\imath)$ is the distance between them, and
$R_\mathrm{tol} = 15$ is the tolerance threshold. The optimal embedding dimension $m^*$ is taken as the
smallest $m$ for which $\mathrm{FNN}(m) < 0.10$ (i.e.\ fewer than 10\% of
neighbours are false). If no such $m$ is found within the search range
$m \in [1, 6]$, the dimension with the minimum FNN fraction is used, with
a floor of $m^* = 2$ to ensure a non-trivial reconstruction.

\subsubsection{Correlation Integral}

Given the $N_v = N - (m^*-1)\tau$ reconstructed vectors $\{\mathbf{y}_i\}$,
the correlation integral counts the fraction of pairs closer than a threshold
$r$:
\begin{equation}
    C(r) = \frac{2}{N_v(N_v-1)}
           \sum_{i<j} \Theta\bigl(r - \|\mathbf{y}_i - \mathbf{y}_j\|\bigr),
\label{eq:CI}
\end{equation}
where $\Theta$ is the Heaviside step function and $\|\cdot\|$ denotes the
Euclidean norm. In practice the full pairwise distance matrix is computed
via \texttt{pdist} and $C(r)$ is evaluated over a logarithmically spaced
grid of 15 radii $r \in [10^{-1.5},\, 10^{0.5}]$.

\subsubsection{Scaling Exponent and $D_2$}

For a self-similar (fractal) attractor, the correlation integral scales as
\begin{equation}
    C(r) \sim r^{D_2} \quad \text{as } r \to 0,
\label{eq:D2scaling}
\end{equation}
so that $D_2$ is the slope of $\log C(r)$ versus $\log r$ in the
scaling region. It is estimated by fitting a straight line to the valid
portion of the log--log curve---specifically those points for which
$0 < C(r) < C_\mathrm{max}$---using ordinary least squares:
\begin{equation}
    D_2 = \frac{d\log C(r)}{d\log r}\bigg|_{\text{scaling region}}.
\label{eq:D2slope}
\end{equation}
An integer value $D_2 = d$ corresponds to a $d$-dimensional torus
(quasi-periodic motion), whereas a non-integer $D_2$ is the hallmark of a
strange attractor and chaotic dynamics. In the present work, $D_2$ is
computed 
alongside the 0--1 test result to provide a consistent characterisation of
the transition between periodic and chaotic flow regimes.

\bibliographystyle{jfm}
\bibliography{jfm}

\end{document}